\documentclass[journal,draftcls,onecolumn,12pt,twoside]{IEEEtran}
\usepackage{pifont}
\usepackage{amsfonts}
\usepackage{amsmath}
\usepackage{mathdots}
\usepackage{bbding}
\usepackage{txfonts}
\usepackage{amssymb}
\usepackage{graphicx}
\usepackage{multirow}
\usepackage{bm}
\usepackage{algorithm}
\usepackage{algpseudocode}
\usepackage{arydshln}

\begin{document}
\title{\footnote{The first revision of this paper according to the comments of the reviewers has been submitted to IEEE Transactions on Communications on June 12, 2019.} Construction of QC LDPC Codes with Low Error Floor by Efficient Systematic Search and Elimination of Trapping Sets}
\author{Bashirreza~Karimi, and Amir.H~Banihashemi, \it{Senior Member, IEEE} 
}

\markboth{}{IEEE Transactions on Communications}

\maketitle

\vspace{-2cm}

\begin{abstract} We propose a systematic design of protograph-based quasi-cyclic (QC) low-density parity-check (LDPC) codes 
with low error floor. We first characterize the trapping sets of such codes and demonstrate that the QC structure of the code eliminates some of the 
trapping set structures that can exist in a code with the same degree distribution and girth but lacking the QC structure. Using this characterization, 
our design aims at eliminating a targeted collection of trapping sets. Considering the parent/child relationship between the trapping sets in the collection, 
we search for and eliminate those trapping sets that are in the collection but are not a child of any other trapping set in the collection. 
An efficient layered algorithm is designed for the search of these targeted trapping sets. 
Compared to the existing codes in the literature, the designed codes are superior in the sense that they are free of the same collection of trapping sets while having a smaller block length, 
or a larger collection of trapping sets while having the same block length. In addition, the efficiency of the search algorithm 
makes it possible to design codes with larger degrees which are free of trapping sets within larger ranges 
compared to the state-of-the-art.   
\end{abstract}

\begin{flushleft}
\end{flushleft}

\vspace{-0.5cm}

\IEEEpeerreviewmaketitle

\section{Introduction} \label{introduction}

Protograph-based quasi-cyclic (QC) low-density parity-check (LDPC) codes are an important category of LDPC codes, adopted in many standards and widely used in practice.
The Tanner graphs for such codes are obtained by cyclically lifting a small bipartite graph, called {\em base graph} or {\em protograph}. Protograph-based QC-LDPC codes not only have a competitive performance under iterative decoding algorithms over a variety of channels, but also enjoy efficient implementations which take advantage of the QC  structure of the code. A potential problem in using QC-LDPC codes in applications that require low error rates is the {\em error floor}, 
characterized by a change in the slope of error rate curves as the channel quality improves.
There are two main approaches to improve the error floor of LDPC codes: (1) modification of the decoding algorithm, and (2) new/modified code constructions. 
This work focusses on the second approach. Increasing the girth of QC-LDPC codes and removing the dominant trapping set structures in the Tanner graph of codes are 
two main methods dealing with the error floor in the construction process of LDPC codes. Progressive-edge-growth (PEG) is one of the most well-known algorithms to design
 LDPC codes with large girth \cite{ref 1}. The PEG algorithm also has been extended to irregular \cite{ref 21} and QC structures \cite{ref 2}. In \cite{ref 4}, 
Asvadi {\em et al.} designed QC-LDPC codes with low error floor by removing the short cycles that were part of dominant trapping sets. The same authors~\cite{ABA-TCOM-2012} also proposed another technique based on the approximate cycle extrinsic message degree (ACE) spectrum to design irregular QC-LDPC codes with good error floor. 
In~\cite{ref 6}, Nguyen \textit{et al.} constructed structured regular LDPC codes with low error floor over the binary symmetric channel (BSC). 
The low error floor in~\cite{ref 6} was achieved by ensuring that certain small trapping sets were absent in the code. Khazraei \textit{et al.} \cite{ref 5} modified 
the PEG algorithm for the construction of LDPC codes to avoid the creation of dominant trapping sets in the construction process. In \cite{ref 7}, Wang \textit{et al.} used the cycle consistency matrix to design separable circulant-based LDPC codes, in which certain absorbing sets were avoided. More recently, 
Diouf \textit{et al.} \cite{ref 8} proposed an improved PEG algorithm to construct regular LDPC codes with variable degree $3$ and girth $8$ 
without $(5,3)$ trapping sets and with the minimum number of $(6,4)$ trapping sets. (An $(a,b)$ trapping set has $a$ variable nodes and $b$ odd-degree check nodes in its subgraph.)
Most recently, in \cite{ref 9}, Tao {\em et al.} proposed a construction of QC-LDPC codes with variable node degree $3$ 
and girth $8$ from fully-connected protographs, where $(a,b)$ elementary trapping sets (ETSs) with 
$a \leq 8$ and $b \leq 3$ were removed by avoiding certain cycles of length $8$ ($8$-cycles) in the Tanner graph. Based on their approach, 
the authors of~\cite{ref 9} also derived lower bounds on the lifting degree, and as a result on the block length, of the designed codes, and were able to find codes, using random 
search, to either achieve or approach the bound.

Elementary trapping sets of LDPC codes are known to be the main culprits in the error floor region over the additive white Gaussian noise (AWGN) channel, see, e.g.,~\cite{ref 10,HB-Arxiv}, and the references therein. 
These sets have been, in general, characterized in \cite{ref 10,HB-Arxiv} using the so-called {\em dpl characterization}.
The $dpl$ characterization describes each and every ETS $S$ as an embedded sequence of ETSs that starts from a simple cycle and expands recursively, in each step by one of the 
three simple expansions $dot$, $path$ and $lollipop$, to reach $S$. Associated with the $dpl$ characterization is an efficient $dpl$ search algorithm~\cite{ref 10,HB-Arxiv}, that can find
all the ETSs within a range of interest, exhaustively. 

In this paper, we first characterize the ETSs of QC-LDPC codes, and demonstrate that some of the ETS structures that can 
exist in a general (randomly constructed) LDPC code are absent in QC-LDPC codes due to the QC structure of the code. To find such structures, we translate 
the problem into an edge coloring problem involving the normal graph \cite{KB-IT-2012} of the ETS structure. Our characterization still follows the $dpl$ principle but
with fewer structures compared to those of a general LDPC code, as considered in \cite{ref 10,HB-Arxiv}.   
Based on the characterization of ETSs in QC-LDPC codes, we then propose a systematic approach to design protograph-based QC-LDPC codes that are free of a certain collection of 
trapping sets. The design is based on investigating the parent/child relationship between all the ETS structures within the collection and target those that are 
not child to any structure within the collection. We then devise an efficient layered algorithm to search for the targeted structures in the construction process. 
Compared to the exhaustive $dpl$ search of~\cite{ref 10,HB-Arxiv}, the proposed search algorithm is significantly less complex. This is mainly due to the fact that while the goal of the $dpl$ algorithm of \cite{ref 10,HB-Arxiv} is to find all the instances of a certain collection of ETS structures, our goal here is only to verify whether any instances of at least one of the 
targeted structures exists in the code. A number of techniques are then employed to solve this new problem efficiently. In particular, the problem
is formulated as a backward recursion in which the goal is to minimize the number of intermediate structures to reach the targeted structures and to use structures that 
have a higher chance of having a smaller multiplicity in the graph. It is important to note that the proposed layered characterization/search algorithm of ETSs can also be used 
to design LDPC codes with low error floor that lack the QC structure. The only difference is that for such codes the number of possible ETS structures is larger.  

The constructed codes in this work are superior to the state-of-the-art codes in the literature in the sense that, with the same protograph, 
they are either free of the same collection of ETSs while having a shorter block length, or are  
free of a larger range of trapping sets (and thus have a superior error floor) while having the same block length. To the best of our knowledge, the proposed design 
is the first that can systematically, efficiently and optimally construct QC-LDPC codes that are free of a certain collection of trapping sets. The systematic (general)
nature of the design means that it is applicable to a variety of QC-LDPC codes with different node degrees, girths and different choices of targeted 
ETS structures. The high efficiency (low complexity) makes it possible to design codes of larger block length with wider range of node degrees and 
to target larger collections of trapping sets. The optimality of the design, which ensures that only the structures of interest are targeted for elimination,
guarantees that the block length is minimized (for the elimination of a given set of ETSs) or that the largest collection of ETSs are removed (for a given block length).
This is in contrast with some of the existing work, such as \cite{ref 8, ref 9}, in which, some parent structures which are not of direct interest are targeted for elimination 
just to eliminate some of their children.

We end this section by noting that there is a body of work that considers absorbing sets to be responsible for the error floor of 
LDPC codes (as an alternative to TSs), see, e.g.,~\cite{ref 7, Lara, SZ, TBF}.  There are close relationships between absorbing sets 
and TSs. In particular, elementary absorbing sets, which are a category of absorbing sets widely studied in the literature~\cite{ref 7,Lara, SZ, TBF}, are in general a subset of leafless ETSs (LETSs)~\cite{ref 10, ref 17}, considered in this work. 
(The two categories of LETSs and elementary absorbing sets are identical for LDPC codes with maximum variable node degree $3$.)  Therefore, the general characterization of LETSs for QC-LDPC codes and  
the layered search algorithm of LETSs proposed in this work are both readily applicable to elementary absorbing sets. We also note that an example of an LDPC code with 
variable node degree $4$ is provided in \cite{ref 10} (${\cal C}_{10}$), for which the error-prone structures of quantized iterative decoders over the AWGN channel in the error floor region 
are shown to be LETSs and not elementary absorbing sets. The existence of such cases further justifies the choice made in this paper to focus on LETSs rather than elementary absorbing sets.

The rest of this paper is organized as follows. In Section II, we present some definitions, notations and preliminaries. The ETS characterization for QC-LDPC codes is 
discussed in Section III. Section IV describes the approach for determining the ETSs within a given collection that are targeted for elimination based on parent/child relationships within the collection, and the proposed layered search algorithm for finding the targeted ETSs efficiently. In Section V, we present the method for the construction 
of QC-LDPC codes with low error floors based on the proposed layered search algorithm. Section VI is devoted to numerical results and presents some of the constructed 
codes and comparisons with existing codes in the literature. The paper is concluded in Section VII.

\section{Preliminaries} 

Consider an undirected bipartite graph $G'(V'= U' \cup W',E')$, where $V'$ and $E'$ are the sets of nodes and edges of $G'$, respectively,
and $U'$ and $W'$ are the sets of nodes on the two sides of the bipartition. Suppose that $|U'|=n$ and $|W'|=m$. 
In this work, we consider bipartite graphs with no parallel edges.  
Suppose the graph ${G}(V=U \cup W,E)$ is constructed from the bipartite graph $G'(V'=U' \cup W',E')$ through the following process: Make $N$ copies of $G'$. Corresponding to every node $v' \in V'$ and every edge $e' \in E'$, generate a set of nodes ${\bf v}=\{v'^0,...,v'^{N-1}\}$ and a set of edges ${\bf e}=\{e'^0,...,e'^{N-1}\}$. Assign a circular permutation $\pi^{e'}$ over the set $\{0,1,\ldots,N-1\}$ to each edge $e'$ in $E'$, and connect the nodes in ${V}$ by the edges in ${E}$ such that if $e'=\{u',w'\}$, for $u' \in U'$ and $w' \in W'$, is in $G'$, then $\{u'^i,w'^j\}$ belongs to ${G}$ if and only if $\pi^{e'}(i)=j$. The graph ${G}$ so constructed is referred to as a {\em cyclic $N$-lifting} of $G'$, with $N$ called the {\em lifting degree}. Graph $G'$, on the other hand, is called the {\em base graph} or {\em protograph}. 

Now, consider an LDPC code whose Tanner graph is ${G}$ with the set of variable and check nodes equal to ${U}$ and ${W}$, respectively. 
Such an LDPC code is a protograph-based QC-LDPC code whose $mN \times nN$ parity-check matrix $H$, given by the bi-adjacency matrix of ${G}$, has the following form: 
\begin{equation}
\footnotesize
\label{eq1}
H=
\begin{bmatrix} 
I^{p_{00}} & I^{p_{01}} & \cdots & I^{p_{0(n-1)}} \\
I^{p_{10}} & I^{p_{11}} & \cdots & I^{p_{1(n-1)}} \\
\vdots & \vdots & \ddots & \vdots \\
I^{p_{(m-1)0}} & I^{p_{(m-1)1}} & \cdots & I^{p_{(m-1)(n-1)}} \\
\end{bmatrix}
\:.
\end{equation}
In (\ref{eq1}), the parameters $p_{ij}$, called {\em permutation shifts}, are in the set $\{0,1,\cdots,N-1,\infty\}$ for $0 \leq i \leq m-1$, $0 \leq j \leq n-1$. The matrix $I^{p_{ij}}$ is obtained by cyclically shifting the rows of the identity matrix $I_{N \times N}$ to the left by $p_{ij}$ units, if $p_{ij} \neq \infty$. Otherwise, $I^{\infty}$ denotes the $N \times N$ all-zero matrix. 
The collection of permutation shifts $p_{ij}$ as a matrix is denoted by $P=[p_{ij}]$ and is called {\em exponent matrix}. 

In this work, we focus on protograph-based QC-LDPC codes, whose base graphs are fully-connected, i.e., every node on each side of the graph is connected to all the nodes on the other side.
This implies that all the nodes in $U'$ have the same degree $d_v = m$ and all the nodes in $W'$ have the same degree $d_c = n$. 
Parameters $d_v$ and $d_c$ are the variable and check node degrees of the constructed regular QC-LDPC code. 
It is well-known that for any QC-LDPC code with a fully-connected base graph, there exists an isomorphic QC-LDPC code with the exponent matrix in the following form (see, e.g.,~\cite{ref 3}):
\begin{equation}
\footnotesize
\label{eq2}
P=
\begin{bmatrix} 
0 & 0 & \cdots & 0 \\
0 & p_{11} & \cdots & p_{1(n-1)} \\
\vdots & \vdots & \ddots & \vdots \\
0 & p_{(m-1)1} & \cdots & p_{(m-1)(n-1)} \\
\end{bmatrix}
\:,
\end{equation}
where $0 \leq p_{11} \leq \cdots \leq p_{1(n-1)}$. 

Given the exponent matrix of a QC-LDPC code, the necessary and sufficient condition for having a $2l$-cycle in the Tanner graph is \cite{ref 3, marc}:
\begin{equation}
\label{eq3}
\sum_{i=0}^{l-1} (p_{m_in_i}-p_{m_in_{i+1}})=0 \text{  mod  } N\:,
\end{equation}
where, $n_0=n_l$, $m_i \neq m_{i+1}$, $n_i \neq n_{i+1}$. The sequence of permutation shifts in (\ref{eq3}) corresponds to a tailless backtrackless closed (TBC) walk \cite{ref 3} 
in the base graph whose permutation shift is equal to the left hand side of (\ref{eq3}). An additional condition for having a $2l$-cycle in the Tanner graph is that the TBC walk 
corresponding to (\ref{eq3}) has no TBC subwalk whose permutation shift is also equal to zero. 
The length of the shortest cycle(s) in a Tanner graph is called {\em girth}, and is denoted by $g$. It is well-known that for good performance $4$-cycles should be avoided and that 
codes with larger girth generally perform better both in waterfall and error floor regions. In our constructions, we impose a lower bound on $g$. When 
the goal is to find a QC-LDPC code with $g \geq g_0$, we choose $p_{ij}$ values in (\ref{eq2}) such that (3) is not satisfied 
for any $l < g_0/2$, and any sequence of $2l$ permutation shifts.

It is well-known that certain substructures of Tanner graphs are responsible for the error floor of LDPC codes. These substructures are
generally referred to as {\em trapping sets}. A trapping set is often characterized by its 
size (the number of variable nodes) $a$ and the number of unsatisfied (odd-degree) check nodes $b$ in its induced subgraph.
Such a trapping set is said to belong to the {\em $(a,b)$ class}. Among trapping sets, {\em elementary trapping sets (ETS)}, 
whose subgraphs have only degree-$1$ and degree-$2$ check nodes, are known to be the most problematic ones~\cite{ref 10},~\cite{HB-Arxiv}. 
Within ETS category, {\em leafless ETSs (LETS)}, those in which each variable node is connected to 
at least two satisfied (even-degree) check nodes, are the most harmful~\cite{ref 10},~\cite{ref 17}. 
Recently, Hashemi and Banihashemi~\cite{ref 10} proposed a hierarchical graph-based characterization of LETS structures, 
dubbed {\em dpl}, for variable-regular LDPC codes.\footnote{An LDPC code is called {\em variable-regular}, if all the variable nodes in the Tanner graph of the code have the same degree.} 
The characterization corresponds to an exhaustive search algorithm 
of LETSs within any range of $a \leq a_{max}$ and $b \leq b_{max}$, for general LDPC codes.

\section{Characterization of ETS Structures in QC-LDPC codes}

In this section, we investigate the constraints that the QC structure of LDPC codes imposes on the ETS structures. We demonstrate that as a result of such constraints, certain ETS structures that can appear in (randomly constructed) LDPC codes, cannot exist in similar QC-LDPC codes. To show this, we transform the problem into a graph coloring problem. 

Consider the induced subgraph $G(S)$ of an ETS $S$ in the Tanner graph $G$ of a variable-regular LDPC code with variable degree $d_v$. 
Replace any degree-$2$ check node and its adjacent edges in $G(S)$ with a single edge, and remove all the degree-$1$ check nodes and their adjacent edges from $G(S)$.
The resulted graph, which is not bipartite, is called the {\em normal graph} of $S$~\cite{KB-IT-2012}, and is denoted by $\hat{G}(S)$ in this paper. There is 
clearly a one-to-one correspondence between $G(S)$ and $\hat{G}(S)$ for variable-regular LDPC codes. To investigate the constraints imposed by the QC structure of the codes, 
in the following, we work with the normal graph representation of ETSs. We note that for Tanner graphs with girth at least $6$, the normal graph of any ETS is simple, i.e., has no parallel edges.

For a graph ${\cal G}$, a {\em $k$-edge coloring} is defined as a function $f : E({\cal G}) \rightarrow C$, such that $|C| = k$, and $f(e) \neq f(e')$ for any two adjacent edges
$e$ and $e'$ of ${\cal G}$. A graph ${\cal G}$ is {\em $k$-edge colorable} if ${\cal G}$ has a $k$-edge coloring. 
The {\em chromatic index} of ${\cal G}$, denoted by $\chi({\cal G})$, is the minimum value of $k$ for which ${\cal G}$ has a $k$-edge coloring. 
It is well-known that for a simple graph ${\cal G}$, $\Delta({\cal G}) \leq \chi({\cal G}) \leq \Delta({\cal G}) + 1$, where $\Delta({\cal G})$ is the maximum node degree of ${\cal G}$~\cite{ref 16}. 
A graph ${\cal G}$ is said to be of Class 1 (resp., Class 2) if $\chi({\cal G}) = \Delta({\cal G})$ (resp., $\chi({\cal G}) = \Delta({\cal G}) + 1$).

{\bfseries{Proposition 1.}} \textit{Consider a QC-LDPC code with the parity-check matrix given by (\ref{eq1}). For an ETS $S$ to exist in the code, 
the normal graph $\hat{G}(S)$ must be $m$-edge colorable.}

\textit{Proof}. It is clear that for protograph-based QC-LDPC codes with the parity-check matrix $H$ of the form (\ref{eq1}), each variable node in an ETS $S$ can be connected 
to at most $m$ check nodes, where each such check node must belong to a distinct row block of $H$. Now, suppose that we assign $m$ different colors to 
different row blocks of $H$, and consider the following assignment of colors to the edges of $\hat{G}(S)$: 
for each edge $e$ in $\hat{G}(S)$, find the row block ${\cal R}$ of $H$ corresponding to the 
degree-$2$ check node in $G(S)$ that has been replaced by $e$. Then, assign the color of ${\cal R}$ to $e$. It is easy to see that for an ETS $S$ to exist in a QC-LDPC code, 
the aforementioned color assignment must be an $m$-edge coloring of $\hat{G}(S)$, or in other words, $\hat{G}(S)$ must be $m$-edge colorable. \hfill $\blacksquare$

We thus have the following result.

{\bfseries{Corollary 1.}} \textit{Consider a QC-LDPC code with the parity-check matrix given by (\ref{eq1}). An ETS $S$ cannot exist 
in the Tanner graph of the code if $\chi(\hat{G}(S)) > m$.}

We note that the result of Corollary 1 is applicable to any QC-LDPC code with the parity-check matrix given by (\ref{eq1}). This includes cases in which some of the 
sub-matrices of (\ref{eq1}) are all-zero, and cases where the degree distribution is irregular. In the latter case, there is no one-to-one correspondence 
between the normal graph $\hat{G}(S)$ of an ETS $S$ and $S$ (see~\cite{HB-Arxiv}, for more information on normal graphs of ETS structures in 
irregular codes and their relationship to quasi-normal hypergraphs of such structures). 

{\bfseries{Corollary 2.}} \textit{An ETS $S$ whose normal graph $\hat{G}(S)$ is of Class 2 with $\Delta(\hat{G}(S)) = m$ cannot exist in 
a QC-LDPC code with $g \geq 6$ and the parity-check matrix given by (\ref{eq1}).}

A graph $G=(V,E)$ is called {\em overfull} if $|E|> \lfloor \frac{|V|}{2} \rfloor \times \Delta(G)$. It is known that an overfull graph is of Class 2~\cite{ref 16}. 
We thus have the following result.

{\bfseries{Proposition 2.}} \textit{Any $(a,b)$ ETS with an odd value of $a$ and with $b < \min(a,d_v)$ cannot exist in a variable-regular QC-LDPC code with variable degree $d_v$ and $g \geq 6$, that is a cyclic lifting of a fully-connected base graph.}

\textit{Proof}. We first note that for an ETS $S$ with $b < a$, we have $\Delta(\hat{G}(S)) = d_v$, because there is at least one variable node in $G(S)$ that is not connected to any 
degree-$1$ check nodes. We then show that under the conditions of the proposition, 
the graph $\hat{G}(S)$ is also overfull and thus of Class 2, This, based on Corollary 2, completes the proof. To show that $\hat{G}(S)$ is overfull, we note that
the number of edges in $\hat{G}(S)$ is $|E|= (ad_v-b)/2$, by the definition of a normal graph. Since $b < d_v$, we have $|E| > (a-1)d_v/2 = \lfloor a/2 \rfloor d_v$,
where the equality is a result of $a$ being odd. \hfill $\blacksquare$

We note that the results of Corollaries 1, 2 and Proposition 2 hold regardless of the lifting degree and permutation shifts of the QC-LDPC code.

Using Proposition 2, we can conclude that some of the ETS classes that can generally exist in (randomly constructed) variable-regular LDPC codes will not appear 
in similar QC-LDPC codes. As an example, consider a variable-regular LDPC code with $d_v=3$ and $g=6$. Based on Table VI of \cite{ref 10}, this code can have 
LETSs in classes $(5,1)$, $(7,1)$ and $(9,1)$. The multiplicity of non-isomorphic structures in these classes are $1$, $4$ and $19$, respectively. Based on Proposition 2,
however, none of these structures can possibly exist in a QC-LDPC code, lifted from a fully-connected base graph, with similar $d_v$ and $g$ values.

It is important to note that while one can use the vast literature on the edge coloring of graphs and derive analytical results 
similar to Proposition 2 for other classes or structures of ETSs, for practical purposes, one can also use an
algorithm for finding the chromatic index of a graph, such as that of \cite{ref 19}, to examine different ETS structures and see 
if their chromatic index is larger than $d_v$ (i.e., it is $d_v+1$). Clearly, a ``yes" answer would mean that such a structure 
cannot exist in a variable-regular QC-LDPC code with variable degree $d_v$. 

In Table~I, we have listed all the classes in which at least one LETS structure does not exist in the QC category of the corresponding variable-regular LDPC codes.
Similar to \cite{ref 10}, in Table~I, the results are separated based on the values of $d_v$ and $g$. For each value of $d_v$ and $g$, and for each class with at least one missing structure, we have provided the multiplicity of the non-isomorphic structures within the class as the bottom entries, where the right and left entries are for the general and the QC cases, respectively. For the entries, we have followed the same notation as in \cite{ref 10}, where the number in the brackets shows the multiplicity of the structures and the notation $s_k$ indicates the simple cycle parent of those structures which has a length of $2k$. As an example, for $d_v=3$ and $g=6$, the bottom entries for the $(5,1)$ class are $s_3(0)/s_3(1)$, which means while a general LDPC code with $d_v=3$ and $g=6$ can have one structure within this class with the parent being a $6$-cycle, this structure cannot exist in the QC category. 
The boldfaced entries in the table highlight the classes for which all the structures are non-existent for the QC category. The non-existence of all these classes follow from Proposition 2.

The non-existence of some LETS structures in QC-LDPC codes compared to their random counterparts not only reduces the search complexity of the proposed technique, as discussed in Subsection~\ref{sec56}.D, but also can be potentially beneficial in achieving a lower error floor.

In the rest of the paper, similar to the existing literature~\cite{TC-2009,TC-2010,ref 10,KB-IT-2012}, we focus on LETSs as the main problematic structures in the error floor. This 
is due to the fact that a vast majority of trapping sets of variable-regular LDPC codes are known to be LETSs~, see, e.g.,~\cite{ref 10}.
We then use the results presented in Table~I to reduce the number of structures that we need to search for in the process of constructing QC-LDPC codes.

\begin{table}[]
\label{tab1}
	\centering
	\tiny
	\caption{LETS classes in which at least one structure is non-existent for LDPC codes with QC structure (notation $s_k(i)$ as an entry of a class means that there are $i$ non-isomorphic structures in the class whose parent is a simple cycle of length $2k$. Right and left entries are for the general and the QC cases, respectively).}
	\label{my-label}
	\begin{tabular}{||c||c|c|c|c|c||}
		\hline
		\multirow{3}{*}{\begin{tabular}{c} 
		$d_v=3$, $g=6$ \end{tabular}} & \begin{tabular}{c}$\mathbf{(5,1)}$ \\ \hdashline $\mathbf{s_3(0)/s_3(1)}$\end{tabular} & 
		                              \begin{tabular}{c}$\mathbf{(7,1)}$ \\ \hdashline $\mathbf{s_3(0)/s_3(3)}$ \\ $\mathbf{s_4(0)/s_4(1)}$ \end{tabular} & 
		                              \begin{tabular}{c}$(8,2)$ \\ \hdashline $s_3(13)/s_3(14)$ \end{tabular} & 
		                              \begin{tabular}{c}$\mathbf{(9,1)}$ \\ \hdashline $\mathbf{s_3(0)/s_3(15)}$ \\ $\mathbf{s_4(0)/s_4(4)}$ \end{tabular} & 
		                              \begin{tabular}{c}$(9,3)$ \\ \hdashline $s_3(42)/s_3(44)$ \\ $s_5(1)/s_5(2)$ \end{tabular} \\ \cline{2-6} 
                                    
                                  & \begin{tabular}{c}$(10,0)$ \\ \hdashline $s_3(12)/s_3(13)$ \\ $s_5(0)/s_5(1)$\end{tabular} & 
                                    \begin{tabular}{c}$(10,2)$ \\ \hdashline $s_3(77)/s_3(85)$ \\ $s_5(0)/s_5(1)$ \end{tabular} & 
                                    \begin{tabular}{c}$(10,4)$ \\ \hdashline $s_3(126)/s_3(129)$ \end{tabular} & 
                                    \begin{tabular}{c}$\mathbf{(11,1)}$ \\ \hdashline $\mathbf{s_3(0)/s_3(91)}$ \\ $\mathbf{s_4(0)/s_4(22)}$ \\ $\mathbf{s_5(0)/s_5(1)}$ \end{tabular} & 
                                    \begin{tabular}{c}$(11,3)$ \\ \hdashline $s_3(337)/s_3(355)$ \\ $s_4(117)/s_4(120)$ \\ $s_5(6)/s_5(7)$ \end{tabular} \\ \cline{2-6} 

		                          & \begin{tabular}{c}$(11,5)$ \\ \hdashline $s_3(324)/s_3(328)$ \end{tabular} & 
		                            \begin{tabular}{c}$(12,0)$ \\ \hdashline $s_3(58)/s_3(63)$  \end{tabular} & 
		                            \begin{tabular}{c}$(12,2)$ \\ \hdashline $s_3(584)/s_3(641)$ \\ $s_4(180)/s_4(184)$ \\ $s_5(8)/s_5(10)$ \end{tabular} & 
		                            \begin{tabular}{c}$(12,4)$ \\ \hdashline $s_3(1279)/s_3(1315)$ \\ $s_4(521)/s_4(524)$ \\ $s_5(50)/s_5(52)$ \end{tabular} & \\ \hline \hline
		                            
		\multirow{2}{*}{\begin{tabular}{c} 
		$d_v=3$, $g=8$ \end{tabular}} & \begin{tabular}{c}$\mathbf{(7,1)}$ \\ \hdashline $\mathbf{s_4(0)/s_4(1)}$ \end{tabular} & 
		                           \begin{tabular}{c}$\mathbf{(9,1)}$ \\ \hdashline $\mathbf{s_4(0)/s_4(4)}$ \end{tabular} & 
		                           \begin{tabular}{c}$(9,3)$ \\ \hdashline $s_5(0)/s_5(1)$ \end{tabular} & 
		                           \begin{tabular}{c}$(10,0)$ \\ \hdashline $s_5(0)/s_5(1)$ \end{tabular} & 
		                           \begin{tabular}{c}$(10,2)$ \\ \hdashline $s_5(0)/s_5(1)$ \end{tabular} \\ \cline{2-6} 
		
		                         & \begin{tabular}{c}$\mathbf{(11,1)}$ \\ \hdashline $\mathbf{s_4(0)/s_4(22)}$ \\ $\mathbf{s_5(0)/s_5(1)}$ \end{tabular} & 
		                           \begin{tabular}{c}$(11,3)$ \\ \hdashline $s_4(114)/s_4(115)$ \\ $s_5(6)/s_5(7)$  \end{tabular} & 
		                           \begin{tabular}{c}$(12,2)$ \\ \hdashline $s_4(178)/s_4(179)$ \\ $s_5(9)/s_5(11)$ \end{tabular} & 
	                               \begin{tabular}{c}$(12,4)$ \\ \hdashline $s_4(479)/s_4(481)$ \\ $s_5(47)/s_5(48)$ \end{tabular} & \\ \hline \hline
	                               
		\multirow{4}{*}{\begin{tabular}{c}
		$d_v=4$, $g=6$ \end{tabular}} & \begin{tabular}{c}$\mathbf{(5,0)}$ \\ \hdashline $\mathbf{s_3(0)/s_3(1)}$ \end{tabular} & 
		                              \begin{tabular}{c}$\mathbf{(5,2)}$ \\ \hdashline $\mathbf{s_3(0)/s_3(1)}$ \end{tabular} & 
		                              \begin{tabular}{c}$(6,2)$ \\ \hdashline $s_3(2)/s_3(3)$ \end{tabular} & 
		                              \begin{tabular}{c}$\mathbf{(7,0)}$ \\ \hdashline $\mathbf{s_3(0)/s_3(2)}$ \end{tabular} & 
		                              \begin{tabular}{c}$\mathbf{(7,2)}$ \\ \hdashline $\mathbf{s_3(0)/s_3(9)}$ \end{tabular} \\ \cline{2-6} 
		                            
		                          & \begin{tabular}{c}$(8,2)$ \\ \hdashline $s_3(27)/s_3(34)$ \end{tabular} & 
		                            \begin{tabular}{c}$(8,4)$ \\ \hdashline $s_3(120)/s_3(122)$ \end{tabular} & 
		                            \begin{tabular}{c}$(8,6)$ \\ \hdashline $s_3(222)/s_3(224)$ \end{tabular} & 
		                            \begin{tabular}{c}$\mathbf{(9,0)}$ \\ \hdashline $\mathbf{s_3(0)/s_3(16)}$ \end{tabular} & 
		                            \begin{tabular}{c}$\mathbf{(9,2)}$ \\ \hdashline $\mathbf{s_3(0)/s_3(152)}$ \\ $\mathbf{s_4(0)/s_4(2)}$ \end{tabular} \\ \cline{2-6}
		                             
		                         & \begin{tabular}{c}$(9,4)$ \\ \hdashline $s_3(642)/s_3(656)$ \end{tabular} & 
		                           \begin{tabular}{c}$(9,6)$ \\ \hdashline $s_3(1352)/s_3(1360)$ \end{tabular} & 
		                           \begin{tabular}{c}$(9,8)$ \\ \hdashline $s_3(1558)/s_3(1561)$ \end{tabular} & 
		                           \begin{tabular}{c}$(10,0)$ \\ \hdashline $s_3(56)/s_3(57)$ \end{tabular} & 
		                           \begin{tabular}{c}$(10,2)$ \\ \hdashline $s_3(709)/s_3(840)$ \end{tabular} \\ \cline{2-6} 
		                         
		                         & \begin{tabular}{c}$(10,4)$ \\ \hdashline $s_3(4106)/s_3(4140)$ \end{tabular} & 
	   	                           \begin{tabular}{c}$(10,6)$ \\ \hdashline $s_3(9334)/s_3(9382)$ \end{tabular} & 
		                           \begin{tabular}{c}$(10,8)$ \\ \hdashline $s_3(11698)/s_3(11719)$ \end{tabular} & 
		                           \begin{tabular}{c}$(10,10)$ \\ \hdashline $s_3(8763)/s_3(8767)$ \end{tabular} & \\ \hline \hline
		                           
		\multirow{1}{*}{\begin{tabular}{c} 
		$d_v=4$, $g=8$ \end{tabular}} & \begin{tabular}{c}\\ $\mathbf{(9,2)}$ \\ \hdashline $\mathbf{s_4(0)/s_4(2)}$ \end{tabular} & 
		                             \begin{tabular}{c}\\ $\mathbf{(11,0)}$ \\ \hdashline $\mathbf{s_4(0)/s_4(2)}$ \end{tabular} & 
		                             \begin{tabular}{c}\\ $\mathbf{(11,2)}$ \\ \hdashline $\mathbf{s_4(0)/s_4(19)}$ \end{tabular} & 
		                             \begin{tabular}{c}\\ $(11,4)$ \\ \hdashline $s_4(163)/s_4(164)$ \end{tabular} & \\ \hline \hline
		                            
		\multirow{2}{*}{\begin{tabular}{c} 
		$d_v=5$, $g=6$ \end{tabular}} & \begin{tabular}{c}$\mathbf{(7,1)}$ \\ \hdashline $\mathbf{s_3(0)/s_3(1)}$ \end{tabular} & 
		                             \begin{tabular}{c}$\mathbf{(7,3)}$ \\ \hdashline $\mathbf{s_3(0)/s_3(6)}$ \end{tabular} & 
		                             \begin{tabular}{c}$(8,2)$ \\ \hdashline $s_3(13)/s_3(16)$ \end{tabular} & 
		                             \begin{tabular}{c}$(8,4)$ \\ \hdashline $s_3(68)/s_3(75)$ \end{tabular} & 
		                             \begin{tabular}{c}$\mathbf{(9,1)}$ \\ \hdashline $\mathbf{s_3(0)/s_3(28)}$ \end{tabular} \\ \cline{2-6} 
		
		                           & \begin{tabular}{c}$\mathbf{(9,3)}$ \\ \hdashline $\mathbf{s_3(0)/s_3(289)}$  \end{tabular} & 
		                             \begin{tabular}{c}$(9,5)$ \\ \hdashline $s_3(1350)/s_3(1356)$  \end{tabular} & 
		                             \begin{tabular}{c}$(9,7)$ \\ \hdashline $s_3(3776)/s_3(3786)$  \end{tabular} & 
		                             \begin{tabular}{c}$(9,11)$ \\ \hdashline $s_3(9526)/s_3(9527)$  \end{tabular} & \\ \hline
		                             		                             
	\end{tabular} 
\end{table}

\section{Efficient Search Algorithm for A Targeted Set of LETSs}
\label{sec56}

The goal of this paper is to construct QC-LDPC codes free of a certain collection ${\cal L}$ of LETS structures.\footnote{The proposed layered characterization/search algorithm of LETSs can also be used 
to construct LDPC codes with low error floor that lack the QC structure. The only difference is that for such codes the number of possible LETS structures is larger.} 
In the literature, this collection is often identified by a certain range of $a$ and $b$ values, i.e., $a \leq a_{max}$ and $b \leq b_{max}$. As we will discuss in Section V, 
to construct the exponent matrix of a QC-LDPC code (for a given lifting degree), a greedy column-by-column search algorithm is 
used to assign the permutation shifts. In the search process, after assigning the permutation values of a new column, one needs 
to check whether the Tanner graph corresponding to the exponent matrix constructed so far contains any instances of the LETS structures within ${\cal L}$.
A naive approach to perform this task would be to use the exhaustive $dpl$ search of~\cite{ref 10} within the specified range, and see if 
the algorithm can find any LETSs within the range. This approach, however, is too complex to use in the construction process that
may require hundreds or even thousands of such searches. Moreover, the information provided by the $dpl$ algorithm of~\cite{ref 10} is 
much more than what we need in the construction process, i.e., the algorithm provides an exhaustive list of all LETSs within the range of interest. 
However, what we need is just to know whether there exists at least one instance of one of the structures of ${\cal L}$ within the graph. 
In this section, we devise an efficient search algorithm for this problem. The efficiency of the algorithm is a result of the following 
considerations: $(a)$ the algorithm only searches for a minimal 
number of structures within ${\cal L}$. These targeted structures, denoted by ${\cal L}_t$, are the ones that are not child to any of the other structures in ${\cal L}$, 
and $(b)$ the algorithm aims to minimize the number of LETS structures that do not belong to ${\cal L}$ but are needed in the 
search for the structures in ${\cal L}_t$. For the choice of these out-of-range structures, the algorithm also gives priority to LETS classes 
that have a higher chance of having smaller multiplicities in the graph. 
Our search algorithm still follows the general framework of $dpl$ characterization/search of \cite{ref 10}, and thus has all the advantages
of the $dpl$ technique. This means that in the proposed search algorithm, each LETS is characterized (and searched for)
as an embedded sequence of LETS structures that starts from a simple cycle and is expanded, at each step, by one of the three expansions $dot$, $path$ and $lollipop$
until it reaches the LETS structure of interest. 
(For a review of these expansions and the corresponding notations, which we closely follow, the reader is referred to~\cite{ref 10}.)

\subsection{Finding the Target LETSs ${\cal L}_t$}

Suppose that we are interested in constructing a QC-LDPC code free of LETSs in the range of $a \leq a_{max}$ and $b \leq b_{max}$. 
To perform this task, we first identify all the non-isomorphic LETS structures within this range based on the values of $d_v$ and $g$, 
and investigate the parent/child relationship between all such structures. (See~\cite{ref 10}.)
We then start from the smallest size of LETSs in the range, i.e., $a = g/2$, and identify all the non-isomorphic $(g/2, b)$ LETS structures with $b \leq b_{max}$. 
These structures are stored in ${\cal L}_t$. We then go through an iterative process, where in each iteration, we increase $a$ by one, until 
we reach $a_{max}$. In each iteration, we examine all the $(a,b)$ LETS structures with the specific $a$ value of that iteration, and $b$ values in the range $b \leq b_{max}$. 
If there is any such structure that is not a child of previously stored structures in ${\cal L}_t$, we add that structure to ${\cal L}_t$. 
By the way that the list ${\cal L}_t$ is constructed, it is clear that it contains the minimum number of structures in ${\cal L}$ that need to be targeted for elimination 
such that none of the LETS structures in ${\cal L}$ can exist in the code. (Elimination of a parent guarantees the elimination of all its children.) 
The following is an example of finding ${\cal L}_t$. 

{\bfseries{Example 1}}. \textit{In this example, we consider a QC-LDPC code construction with $d_v=3$, $g=8$, and four different 
ranges of interest $r_1 : a \leq 6 , b \leq 3$, $r_2 : a \leq 8, b \leq 3$, $r_3 : a \leq 10, b \leq 3$, and $r_4 : a \leq 12, b \leq 3$. 
The parent/child relationships between non-isomorphic LETS structures within the largest range $r_4$ are shown in Fig. 1.
In Fig. 1, the direct (resp., indirect) children are those that are created from their parents by one expansion (resp., multiple expansions).
Different non-isomorphic structures within the same class are identified by different numbers in braces.
From Fig.~1, it is easy to see ${\cal L}_t$ sets for ranges $r_1$ to $r_4$ are ${\cal L}_{t_1} = \{(5,3)\}$, ${\cal L}_{t_2} = {\cal L}_{t_1} \cup \{(7,3)\{1\}, (7,3)\{2\}\}$, 
${\cal L}_{t_3} = {\cal L}_{t_2} \cup \{(9,3)\{7,8,9,10,13,14,15,16,17\}\}$,
and ${\cal L}_{t_4} = {\cal L}_{t_3} \cup \{(11,3)\{45,\ldots,48, 50,\ldots,84, 89, 90, 91, 94, 95, 98,\ldots,108, 112, 117,\ldots,122\}\}$, respectively.
As can be seen, $|{\cal L}_t|$ is considerably smaller than $|{\cal L}|$ for each range. For example for $r_4$, $|{\cal L}_{t_4}|=74$ versus $|{\cal L}|=392$.}

\begin{figure}
	\centering \scalebox{0.55}
	{\includegraphics{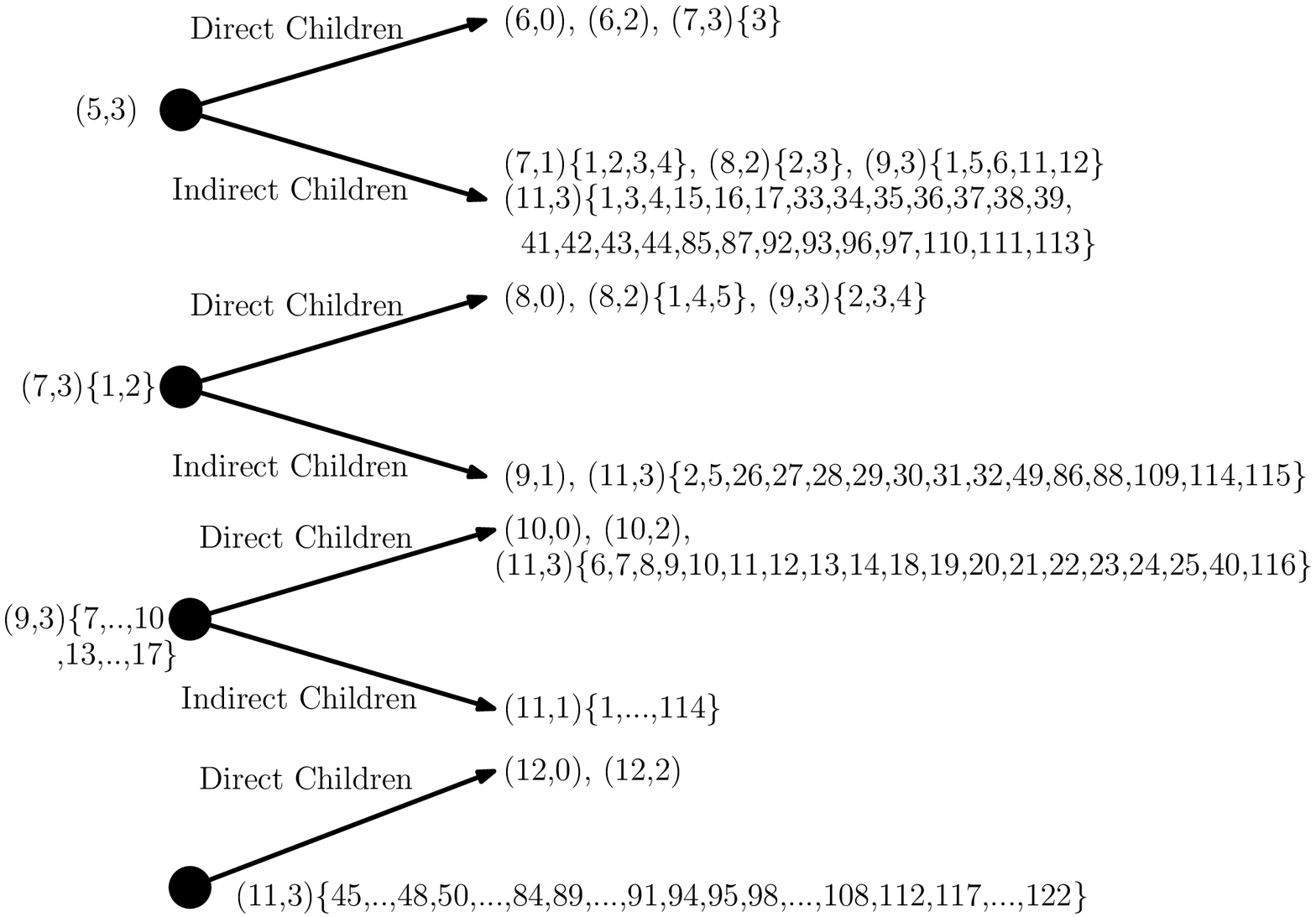}}\\\vspace{.1cm}
	{\normalsize\textbf{\footnotesize Fig. 1}\,\,\,\footnotesize Parent/Child relationships between LETS structures of variable-regular QC-LDPC codes with $d_v=3$, $g=8$, within the range $a \leq 12$ and $b \leq 3$ (different non-isomorphic structures within the same class are identified by different numbers in braces)}\vspace{-1.0cm}
\end{figure}

\subsection{Efficient Search Algorithm for LETSs in ${\cal L}_t$}

To search for all the LETS structures in ${\cal L}_t$, we devise a backward recursion that starts from the LETS structures in  ${\cal L}_t$ with the largest size (largest $a$).
Let such structures be denoted by ${\cal S}_1$, and let ${\cal S}={\cal S}_1$.
Since the structures in ${\cal S}$ cannot be reached through any LETS structure within ${\cal L}_t$ (or ${\cal L}$) using $dpl$ expansions, 
we consider all the possible direct parents ${\cal P}$ of such structures outside ${\cal L}$. The structures in ${\cal P}$ are those that can reach 
at least one of the structures in ${\cal S}$ by the application of just one of the three $dpl$ expansions. We then prioritize the 
classes of the structures in ${\cal P}$ according to certain criteria discussed later. 
Within the class with the highest priority, we then select structures using a greedy iterative process. The process starts by 
selecting the structure in ${\cal P}$ that has the largest number of direct children in ${\cal S}$. Denote this structure by $\xi$. 
We move $\xi$ from ${\cal P}$ to a set $\Pi_1$ ($\Pi_1$ is initially empty), and remove all the direct children of $\xi$ from ${\cal S}$. This ends the first iteration. 
The iterations continue until the set ${\cal S}$ is empty or until no direct parent of the structures in ${\cal S}$ is left in the class with the highest priority.
In the former case, the first step of the recursion is completed. In the latter case, we move on to the class with the second highest priority 
and apply the iterative process. This will continue until the set ${\cal S}$ is empty and thus the first step of the recursion is completed.
At this point, the set $\Pi_1$ contains all the direct parents of the structures in ${\cal S}$.  
To start the next step of recursion, we consider ${\cal S} = \Pi_1\cup {\cal S}_2$ as the new set ${\cal S}$, where ${\cal S}_2$ denotes 
the LETS structures in  ${\cal L}_t$ with the second largest size. The steps of recursion will continue until all the LETSs in ${\cal L}_t$ are covered.
At the end of the recursion, the union of sets $\Pi_i$ contains all the out-of-range LETS structures that need to be included in the search process. A pseudo-code for 
the process of obtaining out-of-range parent structures is given in Algorithm~\ref{tab:search al2}. 

\begin{algorithm}
	\centering
	\scriptsize
	\caption{Finding the out-of-range parent structures needed for the proposed algorithm to search for LETSs within ${\cal L}_t$ }
	\label{tab:search al2}
	\begin{algorithmic}[1]
		\State  \textbf{Input:} ${\cal L}_t$ 
		\Comment Structures in ${\cal L}_t$ belong to classes with sizes $a_1, a_2, \ldots, a_\eta$, such that $a_1<a_2<\hdots<a_\eta$.
		\State  \textbf{Initialization:} ${\cal S}=\emptyset$.
		\For {$k=0,\ldots,\eta-1$}
		\State $\Pi_k \gets \emptyset$.
		\EndFor
		\For {$i=\eta,\ldots,1$}
		\State ${\cal S}_{\eta-i+1} \gets$ LETS structures in ${\cal L}_t$ with size $a_i$.
		\State ${\cal S}={\cal S}_{\eta-i+1} \cup \Pi_{\eta-i}$.
		\State ${\cal P} \gets$ all possible direct parents of structures in ${\cal S}$.
		\Comment For each value of $i$, structures in ${\cal P}$ belong to classes $(a_j^i,b_j^i), j=1, \ldots, \theta_i$, where classes with a smaller index $j$ have a higher priority. 
		\For {$j=1,\ldots,\theta_i$}
		\State {$\Gamma_j \gets$ structures in ${\cal P}$ that are in class $(a_j^i,b_j^i)$.}
		\While {$\Gamma_j \neq \emptyset$}
		\State Choose $\xi \in \Gamma_j$ with the largest number of direct children in ${\cal S}$.
		\State $\Pi_{\eta-i+1} =  \Pi_{\eta-i+1} \cup \{\xi\}$.
		\State Remove $\xi$ from $\Gamma_j$, and remove all the direct children of $\xi$ from ${\cal S}$.
		\If {${\cal S} = \emptyset$}
		\State Break the {\bf for} loop over variable $j$.
		\EndIf
             \EndWhile
		\EndFor
		\EndFor
		\State $\Pi = \Pi_1 \cup \ldots \cup \Pi_{\eta}$.
		\State \textbf{Output:} $\Pi$.
	\end{algorithmic}
\end{algorithm}

Now, we discuss the criteria that we use to prioritize the classes of structures in ${\cal P}$. 
The main idea is to give priority to classes that have a higher chance of having a smaller multiplicity in the graph. 
This translates to a less complex search and smaller memory requirement. Since there is no 
theoretical result available to predict the multiplicity of different LETS classes within a finite Tanner graph, we 
rely on empirical results. In general, experimental results show that, for a given $a$, LETS 
classes with smaller value of $b$ have smaller multiplicity~\cite{ref 10}. The empirical results also show that for two LETS classes $(a,b)$ and $(a',b')$,
where $a < a'$ and $b < b'$, the multiplicity of $(a,b)$ class is generally smaller than that of $(a',b')$ class~\cite{ref 10}. 
Moreover, consider two classes 
$(a,b)$ and $(a',b')$ of structures in ${\cal P}$, where $a < a'$, and let $S$ and $S'$ be two structures in the two classes, respectively.
Also suppose that $S$ and $S'$ are direct parents of two structures $\xi$ and $\xi'$, respectively, where both $\xi$ 
and $\xi'$ are in the same class $(a",b")$ in ${\cal S}$. The following lemma (Lemma $1$) shows that, based on the 
$dpl$ characterization, we must have $b < b'$. (This implies that when all the structures in ${\cal S}$ belong to a single class, 
then one can easily order the direct parent classes of such structures in ${\cal P}$ in accordance with the $a$ or the $b$ value of the classes, with 
classes of smaller $a$ and $b$ values, or the same $a$ value and smaller $b$, having a higher priority in the selection process.)

\textbf{Lemma 1}. \textit{Consider variable-regular LDPC codes with variable degree $d_v \geq 3$ and the $dpl$ characterization of LETS structures within such codes. 
For a given LETS class $(a,b)$, suppose that classes $(a_1,b_1), (a_2,b_2), \ldots, (a_q,b_q)$ are all the direct parent classes of the $(a,b)$ class. 
For any two such parent classes, if $a_i < a_j$, then $b_i < b_j$. } 

\textit{Proof}. From~\cite{ref 10}, one can see that direct parent classes of the $(a,b)$ class can be one of the two following classes: $(a-1,b+2m-d_v))$, 
or $(a-m,b+2-m(d_v-2))$. The first parent class is  for $dot_m$ expansion with $2 \leq m \leq d_v$, and the second parent class is for $pa_m$ expansion with 
$2 \leq m \leq min\{b/(d_v-2),a-g/2-1\}$, and for $lo_m^c$ expansion with $g/2 \leq m \leq min\{(b+1)/(d_v-2),a-g/2-1\}$ and $g/2 \leq c \leq m$.

Since $m \geq 2$, it is clear that the $a$ value for the parent classes corresponding to $pa_m$ and $lo_m^c$ expansions is always smaller than that of the $dot_m$
expansion. So, we first show that the corresponding $b$ values follow the same trend. For this, we prove that the smallest $b$ value for a parent class corresponding to 
$dot_m$ is larger than the largest $b$ value of a parent class corresponding to $pa_m$ or $lo_m^c$. In the case of $dot_m$ expansion, the smallest $b$ value of 
a parent class is $b+2\times2-d_v=b-d_v+4$ since $m \geq 2$, and for $pa_m$ or $lo_m^c$ expansion, the largest $b$ value of a parent class is 
$b+2-2\times(d_v-2)=b-2d_v+6$, which is smaller than $b-d_v+4$ for $d_v > 2$. 

Considering that the parent classes corresponding to $pa_m$ and $lo_m^c$ expansions are identical, to complete the proof, we just need to show that if
$a-m_1 < a - m_2$, for some values of $m_1$ and $m_2$, then $b+2-m_1(d_v-2) < b+2-m_2(d_v-2)$, which is clearly true under the condition that $d_v > 2$.
\hfill $\blacksquare$

Based on the above discussions, to prioritize the structures in ${\cal P}$, we first prioritize the classes: classes with smaller $a$ and $b$ values have 
a higher priority and for the same value of $a$, classes with smaller $b$ values have a higher priority. Within each class, we then prioritize the structures 
based on the number of direct children that they have in ${\cal S}$, where larger number of children means higher priority. 

\textbf{Example 2}. Consider the code construction discussed in Example~1, where all the LETSs in $r_4$ are to be avoided. 
In this case, the set ${\cal S}_1$ consists of $62$ structures in the $(11,3)$ class. The careful study of these structures reveals that
all the (out-of-range) direct parents of these structures belong to $(10,4)$ and $(10,6)$ classes, and that the structures in 
${\cal S}_1$ are generated by the application of $dot_2$ and $dot_3$ expansions to these parent structures. Between the two classes 
of $(10,4)$ and $(10,6)$, the priority is given to the parent structures in the $(10,4)$ class since for the same value of $a$, this 
class has a smaller $b$ value compared to the $(10,6)$ class. In total, $63$ structures exist in the $(10,4)$ class. By prioritizing these 
structures based on the number of their direct children and choosing them iteratively in the order of their priority, we can cover  
all the structures in ${\cal S}_1$ by choosing only $22$ structures in the $(10,4)$ class. These structures can generate 
all the structures in ${\cal S}_1$ with the $dot_2$ expansion. This ends the first step of the recursion. (see the last level of the trellis diagram in Fig. 2. To prevent the figure from being over crowded, rather than making a connection between each parent/child pair, for the last level, we have shown parents and their children collectively.)

For the second step, we have ${\cal S}= \Pi_1 \cup {\cal S}_2$, where $\Pi_1$ is the set of the $22$ structures in the $(10,4)$ class and ${\cal S}_2$ consists of the $9$ 
LETS structures in the $(9,3)$ class of ${\cal L}_t$. Investigating the direct (out-of-range) parent structures for the set ${\cal S}_2$, we find that they all belong to $(8,4)$ and $(8,6)$ classes, while for the $22$ structures in the $(10,4)$ class, the direct parent structures are in $(8,4)$, $(9,5)$, and $(9,7)$ classes.  
Among the direct parent classes of the set ${\cal S}$, the highest priority is given to the $(8,4)$ class since it has the smallest $a$ and $b$ values. 
By the application of the iterative selection process to the structures in the $(8,4)$ class, we can generate all the $9$ structures in the $(9,3)$ class and $7$ out of $22$ structures 
in the $(10,4)$ class, by using only $7$ structures in the $(8,4)$ class. The remaining $15$ structures in the $(10,4)$ class, however, cannot be generated by
the structures in the $(8,4)$ class. To generate these remaining structures, the parent class with the highest priority (out of the two remaining classes that are direct parent classes 
of the $(10,4)$ class) is the $(9,5)$ class. By the application of the iterative process to this class, the remaining $15$ structures of the $(10,4)$ class are generated through only $5$
structures in the $(9.5)$ class. (See the second last level of the trellis diagram in Fig. 2.)

In the next step of the recursion, the $7$ and $5$ structures from the $(8,4)$ and $(9,5)$ classes, respectively, along with the two $(7,3)$ structures 
in ${\cal L}_t$ form the set ${\cal S}$. The outcome of this step of recursion can be seen in Fig. 2, together with the final step of the recursion, 
that has all the simple cycles of length $8$ and $10$ as the output set $\Pi_4$. 

\begin{figure}
	\centering \scalebox{0.42}
	{\includegraphics{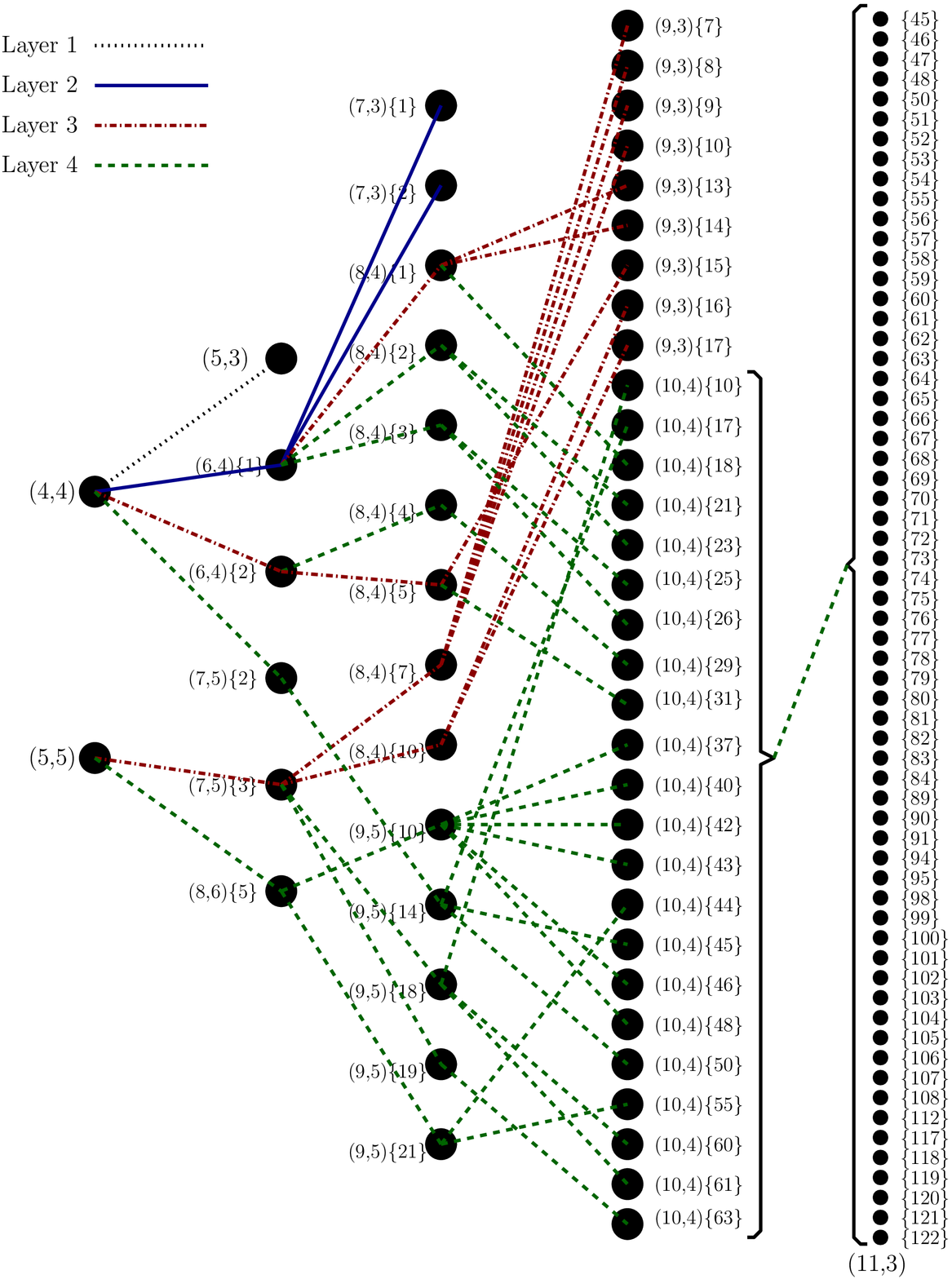}}\\\vspace{.1cm}
	{\normalsize\textbf{\footnotesize Fig. 2}\,\,\,\footnotesize Search tree (trellis diagram) corresponding to LETS structures in the 
        range $a \leq 12$ and $b \leq 3$ for QC-LDPC codes with $d_v=3$ and $g=8$}\vspace{-0.7cm}
\end{figure}

Corresponding to the trellis diagram of Fig.~2, we have the characterization table of LETSs for the proposed search, given at the top of Table II, 
in comparison with the characterization table corresponding to the exhaustive search of~\cite{ref 10} (limited to structures that exist in QC-LDPC codes, as explained in Section III), 
given at the bottom. The entries of the table
correspond to different classes of LETSs. For each class, the top entry shows the non-isomorphic structures within the class that are involved 
in the search process and the bottom entry shows the expansions that will have to be applied to all the instances of those structures.  
By comparing the entries for similar classes in the top and the bottom parts of Table II, one can see 
that both the number of structures and the variety of required expansions have decreased substantially in the proposed search algorithm 
compared to the exhaustive search of~\cite{ref 10}. As an example, for the $(9,5)$ class, the proposed search algorithm only 
needs to apply $dot_2$ expansion to $5$ structures. In comparison, the exhaustive search algorithm of \cite{ref 10} 
requires the application of $dot_2,dot_3$ and $pa_2$ expansions to $20$ structures.

\begin{table*}[]
	\centering
	\tiny
	\caption{Characterization tables of LETSs in QC-LDPC codes with $d_v=3$, $g=8$, within the range $a \leq 12$ and $b \leq 3$: 
        proposed search algorithm (top) and exhaustive search algorithm of \cite{ref 10} (bottom)}
	\label{my-label}
	\begin{tabular}{||c|c|c|c|c|c|c|c|c|c||}
		\hline
		Proposed & $a=4$ & $a=5$ & $a=6$ & $a=7$ & $a=8$ & $a=9$ & $a=10$ & $a=11$ & $a=12$ \\ \hline
        
        $b=0$ & $-$ & $-$ & $-$ & $-$ & $-$ & $-$ & $-$ & $-$ & $-$ \\ \hline

        $b=1$ & $-$ & $-$ & $-$ & $-$ & $-$ & $-$ & $-$ & $-$ & $-$ \\ \hline
        
        $b=2$ & $-$ & $-$ & $-$ & $-$ & $-$ & $-$ & $-$ & $-$ & $-$ \\ \hline
        
        $b=3$ & $-$ & \begin{tabular}{c}$s_4(1)$ \\ \hdashline $-$ \end{tabular} & $-$ & \begin{tabular}{c}$s_4(2)$ \\ \hdashline $-$ \end{tabular} & $-$ & \begin{tabular}{c}$s_4(3),s_5(6)$ \\ \hdashline $-$ \end{tabular} & $-$ & \begin{tabular}{c}$s_4(52)$ \\ $s_5(10)$ \\ \hdashline $-$ \end{tabular} & $-$ \\ \hline
        
        $b=4$ & \begin{tabular}{c}$s_4(1)$ \\ \hdashline $dot_2,pa_2$ \\ $pa_3$  \end{tabular} & $-$ & \begin{tabular}{c}$s_4(2)$ \\ \hdashline $dot_2,pa_2$  \end{tabular}  & $-$ & \begin{tabular}{c}$s_4(5)$,$s_5(2)$ \\ \hdashline $dot_2,pa_2$ \end{tabular} & $-$ & \begin{tabular}{c}$s_4(9)$,$s_5(13)$ \\ \hdashline $dot_2$ \end{tabular} & $-$ & $-$ \\ \hline
         
        $b=5$ & $-$ & \begin{tabular}{c}$s_5(1)$ \\ \hdashline $pa_2,pa_3$  \end{tabular} & $-$ & \begin{tabular}{c}$s_4(1),s_5(1)$ \\ \hdashline $dot_2,pa_2$  \end{tabular} & $-$ & \begin{tabular}{c}$s_4(1),s_5(4)$ \\ \hdashline $dot_2$  \end{tabular} & $-$ & $-$ & $-$ \\ \hline
        
        $b=6$ & $-$ & $-$ & $-$ & $-$ & \begin{tabular}{c}$s_5(1)$ \\ \hdashline $dot_2$  \end{tabular} & $-$ & $-$ & $-$ & $-$ \\ \hline \hline 
        
        \cite{ref 10} & $a=4$ & $a=5$ & $a=6$ & $a=7$ & $a=8$ & $a=9$ & $a=10$ & $a=11$ & $a=12$ \\ \hline
        
        $b=0$ & $-$ & $-$ & \begin{tabular}{c}$s_4(1)$ \\ \hdashline $-$  \end{tabular} & $-$ & \begin{tabular}{c}$s_4(2)$ \\ \hdashline $-$  \end{tabular} & $-$ & \begin{tabular}{c}$s_4(5)$ \\ \hdashline $-$  \end{tabular} & $-$ & \begin{tabular}{c}$s_4(20)$ \\ $s_5(2)$ \\ \hdashline $-$  \end{tabular} \\ \hline
        
        $b=1$ & $-$ & $-$ & $-$ & $-$ & $-$ & $-$ & $-$ & $-$ & $-$ \\ \hline
        
        $b=2$ & $-$ & $-$ & \begin{tabular}{c}$s_4(1)$ \\ \hdashline $pa_3$ \\ $pa_4,lo^4_4$ \end{tabular} & $-$ & \begin{tabular}{c}$s_4(5)$ \\ \hdashline $pa_3$ \end{tabular} & $-$ & \begin{tabular}{c}$s_4(27)$ \\ \hdashline $pa_2$ \end{tabular} & $-$ & \begin{tabular}{c}$s_4(178)$\\$s_5(9)$ \\ \hdashline $-$ \end{tabular} \\ \hline
        
        $b=3$ & $-$ & \begin{tabular}{c}$s_4(1)$ \\ \hdashline $dot_2,dot_3$ \\ $pa_3,pa_4,lo^4_4$ \end{tabular} & $-$ & \begin{tabular}{c}$s_4(3)$ \\ \hdashline $dot_2,dot_3$ \\ $pa_2,pa_3$ \end{tabular} & $-$ & \begin{tabular}{c}$s_4(16)$ \\ \hdashline $dot_2,dot_3$ \\ $pa_2$ \end{tabular} & $-$ & \begin{tabular}{c}$s_4(115)$\\$s_5(7)$ \\ \hdashline $dot_2,dot_3$ \end{tabular} & $-$ \\ \hline
        
        $b=4$ & \begin{tabular}{c}$s_4(1)$ \\ \hdashline $dot_2,pa_2$ \\ $pa_3$  \end{tabular} & $-$ & \begin{tabular}{c}$s_4(2)$ \\ \hdashline $dot_2,dot_3$ \\ $pa_2,pa_3$  \end{tabular}  & $-$ & \begin{tabular}{c}$s_4(9)$,$s_5(1)$ \\ \hdashline $dot_2,dot_3$ \\ $pa_2$ \end{tabular} & $-$ & \begin{tabular}{c}$s_4(57)$,$s_5(6)$ \\ \hdashline $dot_2,dot_3$ \end{tabular} & $-$ & $-$ \\ \hline
        
        $b=5$ & $-$ & \begin{tabular}{c}$s_5(1)$ \\ \hdashline $dot_2,dot_3$ \\ $pa_2,pa_3$  \end{tabular} & $-$ & \begin{tabular}{c}$s_4(2),s_5(1)$ \\ \hdashline $dot_2,dot_3$ \\ $pa_2$  \end{tabular} & $-$ & \begin{tabular}{c}$s_4(16),s_5(4)$ \\ \hdashline $dot_2,dot_3$ \\ $pa_2$  \end{tabular} & $-$ & $-$ & $-$ \\ \hline
        
        $b=6$ & $-$ & $-$ & $-$ & $-$ & \begin{tabular}{c}$s_5(2)$ \\ \hdashline $dot_2,dot_3$  \end{tabular} & $-$ & $-$ & $-$ & $-$ \\ \hline
		
	\end{tabular}
	\vspace{-1.0cm}
\end{table*}

\subsection{Layering of the Search Algorithm}

To further reduce the complexity and memory requirement of our proposed search algorithm, we implement the algorithm in multiple layers.
We recall that the purpose of our search is to determine whether there exists at least one instance of one of the structures of ${\cal L}$ 
in the graph. Consider the structures in ${\cal L}_t$, and assume that they belong to classes with size $a_1, a_2, \ldots, a_\eta$, where $a_1<a_2<\hdots<a_\eta$.
To make the determination that whether there exists at least one instance of one of the structures of ${\cal L}$ 
in the graph, rather than searching for all the structures in ${\cal L}_t$ in one shot, we first search for the structures in the class with size $a_1$ (Layer $1$). 
If we can find at least one instance of a structure in this class, we terminate the search with a positive response. If the graph has 
none of the structures of ${\cal L}_t$ with size $a_1$, we continue our search to Layer $2$, which contains structures in ${\cal L}_t$ with size $a_2$.
We continue this process, until we either find one instance of one structure within one of the layers, or finish the search through all layers without finding 
any LETS in ${\cal L}_t$. In the former case, the search is terminated with a positive response, while in the latter case, the response is negative.

As an example, the four layers of the proposed search algorithm for the construction discussed in Examples $1$ and $2$, corresponding to classes
$(5,3)$, $(7,3)$, $(9,3)$ and $(11,3)$ in ${\cal L}_t$ are identified by different line types in Fig. 2. 

\textbf{Remark 1}. \textit{For further memory reduction, in each layer of the search algorithm, we only keep those LETSs that are needed 
for next layer(s) of our search algorithm. For example, in the second layer of Fig. 2 (solid lines), we only need to 
keep $(6,4)\{1\}$ LETSs since these trapping sets are needed in third and fourth layers. Moreover, to further reduce the complexity,
we can perform the search process within each layer sequentially. For example, in the third layer of Fig. 2 (dash-dot lines), 
we can first apply the expansions corresponding to the $(9,3)$ structures originated from cycles of length $8$, 
and in the case that these trapping sets were missing in the graph, we can continue to search for the rest of the structures in the $(9,3)$ 
class that are originated from cycles of length 10.}

\subsection{Complexity of the Search Algorithm}

In general, the complexity of the search algorithm depends on the multiplicity of different LETS structures involved in the search of the graph 
and the expansions that are applied to them. There is however no theoretical result for the multiplicity of different LETS structures in finite graphs. 
Moreover, as explained before, the graph itself changes throughout the construction process. In addition, in the layered implementation of the algorithm, 
the complexity of the search can highly vary depending on the layer at which an instance of a structure may be found by the algorithm.   
For these reasons, it is difficult to evaluate the complexity of the proposed search algorithm theoretically.

To compare the complexity of the proposed algorithm with that of the exhaustive search algorithm of \cite{ref 10}, however, we count
the number of different LETS structures involved in the search and the different expansions that are applied to them. For this, we use 
characterization tables such as those of Table II, and calculate the weighted sum of each expansion over different entries of the table 
with the weights being the multiplicity of the structures in the corresponding class. Since different expansions have different complexities~\cite{ref 10}, 
we count the weighted sum for each expansion, separately. These results for three construction scenarios are listed in Table III:
$d_v=3, g=8, r: a \leq 12, b \leq 3$; $d_v=4,g=6, r: a \leq 8, b \leq 5$; and $d_v=3,g=6, r: a \leq 11, b \leq 2$. For each scenario,
we have listed the results for four search algorithms in four columns. The first column corresponds to the exhaustive search algorithm of \cite{ref 10} 
where the QC structure of the graph is not taken into account. These results correspond to the characterization tables reported in~\cite{ref 10}. 
In the second column, labeled as ``QC,'' we have reported the results corresponding to the exhaustive search algorithm of \cite{ref 10}, 
where the QC structure of the graph is taken into account. The difference between this case and the first was explained in Section III (Table I).
Finally, the third and fourth columns correspond to the proposed search algorithm for a general and a QC graph, respectively (where the advantages of layering are ignored). 
As an example of how the results in Table III are obtained, consider the first construction scenario and the expansion $pa_3$. For this case, the results reported in 
Table III for the proposed QC search and QC search are $2$ and $14$, respectively. By examining the top part of Table II, one can see that
$pa_3$ appears in only two entries corresponding to classes $(4,4)$ and $(5,5)$, where each class has only one structure, hence the result $1+1=2$.
In the bottom table, however, $pa_3$ appears in $7$ entries corresponding to classes $(4,4), (5,3), (5,5), (6,2), (6,4), (7,3)$, and $(8,2)$, where
each class has the following number of structures: $1, 1, 1, 1, 2, 3$, and $5$, respectively, which add up to $14$.  

The comparison of the results presented in Table III shows the considerable advantage of the proposed algorithm over the exhaustive search 
of \cite{ref 10}, for both a general and a QC graph. One should also note that further advantage is gained through the layered implementation 
of the proposed search algorithm. Table III also demonstrates that the difference between the complexity of searching a general graph and that of searching a QC graph 
is smaller for the proposed algorithm compared to the exhaustive search algorithm of \cite{ref 10}.  

\begin{table}[]
	\centering
	\tiny
	\caption{Complexity comparison between four search algorithms of LETSs: exhaustive search algorithm of \cite{ref 10} for a general and a QC graph, and the 
	proposed search algorithm for a general and a QC graph}
	\label{my-label}
	\begin{tabular}{||c|c|c|c|c|||c|c|c|c|||c|c|c|c||}
		\hline
		&\multicolumn{4}{c|||}{$d_v=3,g=8$}&\multicolumn{4}{c|||}{$d_v=4,g=6$}&\multicolumn{4}{c||}{$d_v=3,g=6$}\\
		&\multicolumn{4}{c|||}{$a \leq 12, b \leq 3$}&\multicolumn{4}{c|||}{$a \leq 8, b \leq 5$}&\multicolumn{4}{c||}{$a \leq 11, b \leq 2$}\\
		\cline{2-13}		
		& \cite{ref 10} & QC & Proposed & Proposed QC & \cite{ref 10} & QC & Proposed & Proposed QC & \cite{ref 10} & QC & Proposed & Proposed QC \\ \hline				
		$dot_2$ & 279 & 244 & 42 & 40 & 70 & 67 & 3 & 3 & 132 & 108 & 62 & 59  \\ \hline
		$dot_3$ & 278 & 243 & 0 & 0 & 124 & 103 & 30 & 29 & 108 & 73 & 46 & 44  \\ \hline
		$dot_4$ & $-$ & $-$ & $-$ & $-$ & 110 & 90 & 1 & 1 & $-$ & $-$ & $-$ & $-$  \\ \hline
		$pa_2$ & 85 & 83 & 14 & 13 & 5 & 4 & 1 & 1 & 40 & 39 & 11 & 11  \\ \hline
		$pa_3$ & 14 & 14 & 2 & 2 & $-$ & $-$ & $-$ & $-$ & 8 & 8 & 3 & 3 \\ \hline
		$pa_4$ & 2 & 2 & $-$ & $-$ & $-$ & $-$ & $-$ & $-$ & $-$ & $-$ & $-$ & $-$  \\ \hline
		$lo^3_3$ & $-$ & $-$ & $-$ & $-$ & $-$ & $-$ & $-$ & $-$ & 5 & 1 & $-$ & $-$  \\ \hline
		$lo^3_4$ & $-$ & $-$ & $-$ & $-$ & $-$ & $-$ & $-$ & $-$ & 2 & 1 & 1 & 1  \\ \hline
		$lo^4_4$ & 3 & 2 & $-$ & $-$ & $-$ & $-$ & $-$ & $-$ & 2 & 1 & $-$ & $-$  \\ \hline	
	\end{tabular}
	\vspace{-1.0cm}
\end{table}

\section{Construction of QC-LDPC Codes with Low Error Floor}

In this part, we propose a construction method for protograph-based QC-LDPC codes with low error floor. The low error floor is achieved by avoiding the LETS structures within a 
predetermined list of structures ${\cal L}$. This is performed using the proposed search algorithm discussed in Section IV.

In this work, we tackle two problems related to designing QC-LDPC codes: (a) For a given base graph of size $m \times n$, a given even integer $g_0$, 
and a given range $a \leq a_{max}$ and $b \leq b_{max}$, find a code with girth at least $g_0$ that has the minimum lifting degree (block length) and does not have 
any $(a,b)$ LETS with $a \leq a_{max}$ and $b \leq b_{max}$; (b) For a given base graph, a given even integer $g_0$, a given positive integer $b_{max}$, 
and a fixed lifting degree $N$, find a code with girth at least $g_0$ and lifting degree $N$ that has no $(a,b)$ LETS in the range $a \leq a_{max}$ and $b \leq b_{max}$, 
where $a_{max}$ is maximized. We note that formulation of Problem (a) is similar to the formulation used in~\cite{ref 9}. A similar formulation is also 
commonly used in the context of designing QC-LDPC codes of a certain girth with minimum length (see, e.g.,~\cite{ref 3}). We further note that in Problem (a), 
rather than a range for $a$ and $b$ values, one can use a list ${\cal L}$ of LETS structures to be avoided.

To design a QC-LDPC code based on a given base graph and a given lifting degree, we need to determine the nonzero elements of the 
exponent matrix of (2). The design constraints are to maintain the girth to at least $g_0$, and to ensure that no target LETS structure 
from the list ${\cal L}$ exists in the code. In this work, to design the code, we use a greedy search algorithm in which the nonzero elements 
of the exponent matrix $P$ are selected column by column starting from the leftmost column. At each step, all the nonzero 
elements of one column of $P$ are assigned in random simultaneously. The sub-matrix of the parity-check matrix $H$ corresponding to the selected columns of $P$ 
so far is then searched to see if the girth constraint is satisfied (no cycle of length less than $g_0$ exists in the corresponding graph) 
and that none of the structures in ${\cal L}$ exists in the sub-matrix. If the sub-matrix satisfies these constraints, 
the algorithm moves on to the next step and assigns the permutation shifts of the next column of $P$. If the selected column fails 
to satisfy the constraints, a new random column is selected and tested. This process will continue until a column is found that 
satisfies the constraints or all the possible choices for that specific column are exhausted (although the choices are made randomly, repeated choices are avoided). 
In the latter case, the algorithm back tracks to the previously selected column and makes a new random choice for it. 
The algorithm will continue until all the columns of $P$ are assigned and the corresponding $H$ matrix satisfies all the constraints, 
or until all the possibilities are exhausted without reaching a solution. A third possibility would be to stop the algorithm if 
it fails to find a solution within a predetermined time period. Note that to search for the LETSs (of ${\cal L}$) in the graph 
corresponding to the sub-matrix of $H$ at each step of the algorithm, we use the proposed layered search algorithm of Section IV.

\section{Numerical Results}

In this section, we report some of the constructed QC-LDPC codes based on the technique described in Section V. All the simulation results
are for binary-input AWGN channel, and a $5$-bit min-sum decoder with clipping threshold equal to 2~\cite{ref 11}, 
and maximum number of iterations 100. For each simulated point, at least $100$ block errors are collected.

As the first experiment, we tackle Problem (a) in Section V for a fully-connected $3 \times 5$ base graph ($d_v=3, d_c=5$), with the constraint $g_0=8$, and with 
no LETSs in ranges $r_1$ to $r_4$, described in Example 1. The lifting degrees and the exponent matrices of the constructed codes are given in 
Table IV. To obtain the results for $r_1: a \leq 6,b \leq 3$, we start from the minimum lifting degree required for having a code with girth $8$, which is $N=13$~\cite{ref 3}, 
and increase the lifting degree $N$ by one at each step until we can find a code which is free of LETSs within the range of interest. 

For $r_1$, the result presented in Table IV is optimal in that $N=18$ is the smallest lifting degree that can result in no LETSs within $r_1$,
(i.e., all the possible exponent matrices for all the values of $13 \leq N \leq 17$ were checked and none was completely free of LETSs in $r_1$). 
To obtain the results for the other ranges, since $r_1 \subset r_2 \subset r_3 \subset r_4$, for each range $r_i, 2 \leq i \leq 4$, we start from the 
smallest $N$ value obtained for $r_{i-1}$. For ranges $r_2, r_3, r_4$, however, 
the value of $N$ presented in Table IV is an upper bound on the smallest lifting degree satisfying the design constraints, i.e., as the value of $N$ was increased at each step by one, 
we limited the search time for each $N$ to $3$ hours.\footnote{Our search algorithm is implemented in MATLAB and is run on a PC with 3.50 GHZ CPU and $32$ GB RAM.} If no exponent 
matrix satisfying all the constraints was found during this time, we would increase $N$ by one and restart the search. 

\begin{table}[]
	\centering
	\tiny
	\caption{Upper bounds on the lifting degree of girth-$8$ QC-LDPC codes with fully-connected $3 \times 5$ base graph with no 
         LETSs within different ranges, and the corresponding exponent matrices}
	\label{my-label}
	\begin{tabular}{||c|c|c|c|c||}
		\hline
		Range & $(a \leq 6,b \leq 3)$ & $(a \leq 8,b \leq 3)$ & $(a \leq 10,b \leq 3)$ & $(a \leq 12,b \leq 3)$ \\ \hline
		$N$ & 18 & 26 & 36 & 46 \\ \hline
		Exponent Matrix & $\begin{bmatrix} 0 & 0 & 0 & 0 & 0 \\ 0 & 1 & 3 & 7 & 8 \\ 0 & 2 & 11 & 5 & 14 \\ \end{bmatrix}$ & $\begin{bmatrix} 0 & 0 & 0 & 0 & 0 \\ 0 & 1 & 2 & 6 & 16 \\ 0 & 3 & 21 & 12 & 23 \\ \end{bmatrix}$ & $\begin{bmatrix} 0 & 0 & 0 & 0 & 0 \\ 0 & 1 & 14 & 29 & 34 \\ 0 & 2 & 24 & 32 & 12 \\ \end{bmatrix}$ & $\begin{bmatrix} 0 & 0 & 0 & 0 & 0 \\ 0 & 1 & 12 & 28 & 33 \\ 0 & 2 & 22 & 35 & 39 \\ \end{bmatrix}$ \\ \hline
	\end{tabular}
	\vspace{-1.0cm}
\end{table}

Similar approach was used for the $3 \times 6$ base graph to design rate-$1/2$ QC-LDPC codes with girth $8$ and free of LETSs within ranges $r_1$ to $r_4$.
The results are presented in Table V.

\begin{table}[p]
	\centering
	\tiny
	\caption{Upper bounds on the lifting degree of girth-$8$ QC-LDPC codes with fully-connected $3 \times 6$ base graph with no 
         LETSs within different ranges, and the corresponding exponent matrices}
	\label{my-label}
	\begin{tabular}{||c|c|c|c|c||}
		\hline
		Range & $(a \leq 6,b \leq 3)$ & $(a \leq 8,b \leq 3)$ & $(a \leq 10,b \leq 3)$ & $(a \leq 12,b \leq 3)$ \\ \hline
		$N$ & 32 & 41 & 60 & 80 \\ \hline
		Exponent Matrix & \tiny{$\begin{bmatrix} 0 & 0 & 0 & 0 & 0 & 0 \\ 0 & 4 & 11 & 17 & 24 & 29 \\ 0 & 14 & 30 & 5 & 3 & 6 \\ \end{bmatrix}$} & \tiny{$\begin{bmatrix} 0 & 0 & 0 & 0 & 0 & 0 \\ 0 & 16 & 18 & 19 & 22 & 33 \\ 0 & 32 & 21 & 26 & 39 & 1 \\ \end{bmatrix}$} & \tiny{$\begin{bmatrix} 0 & 0 & 0 & 0 & 0 & 0 \\ 0 & 23 & 33 & 38 & 40 & 59 \\ 0 & 22 & 45 & 54 & 48 & 42 \\ \end{bmatrix}$} & \tiny{$\begin{bmatrix} 0 & 0 & 0 & 0 & 0 & 0 \\ 0 & 7 & 39 & 41 & 45 & 61 \\ 0 & 35 & 43 & 51 & 66 & 36 \\ \end{bmatrix}$} \\ \hline
	\end{tabular}
\end{table}

In \cite{ref 9}, Problem (a) was tackled for fully-connected base graphs of size $3 \times 5$ and $3 \times 6$, $g=8$ and $r_2$, i.e., $a \leq 8,b \leq 3$.
Based on the design approach of~\cite{ref 9}, it was shown that the minimum lifting degrees required to satisfy the trapping set constraint for 
the two base graphs are $N=41$ and $N=61$, respectively. Using a search algorithm, the authors of~\cite{ref 9} were able to find 
QC-LDPC codes with girth $8$ and lifting degrees $41$ and $63$ for the two base graphs, respectively, that were free of LETSs within $r_2$, 
In comparison, the codes designed here have $N=26$ and $N=41$ for $3 \times 5$ and $3 \times 6$ base graphs, respectively. 
These codes are significantly superior to those found in~\cite{ref 9}, in the sense that they have the same girth and 
degree distribution and satisfy the same trapping set constraint but have a much smaller block length. 
The main reason for the improved results in this work compared to \cite{ref 9} is that, contrary to the approach adopted here which in fact 
imposes a necessary and sufficient condition for removing all the LETS structures within the targeted range, in \cite{ref 9}, 
the authors targeted the $(5,3)$ and $(6,4)\{1\}$ structures. Note that $(6,4)\{1\}$ structure is a parent to some of the targeted structures in the range 
but is not of interest itself. This imposes a sufficient but not necessary condition for removing the targeted LETSs. 
This unnecessary constraint imposed on the design degrades the quality of the achievable solution. 

\begin{table}[]
	\centering
	\tiny
	\caption{Multiplicities of LETS structures in the range $a \leq 12$ and $b \leq 4$ for Code ${\cal C}_1$ and the code designed in~\cite{ref 9}}
	\label{my-label}
	\begin{tabular}{||c|c|c||c|c|c||}
		\hline
		$(a,b)$ class & ${\cal C}_1$ & Code of \cite{ref 9} & $(a,b)$ class & ${\cal C}_1$ & Code of \cite{ref 9} \\ \hline		
		$(4,4)$ & 451 & 451 & $(10,4)$ & 8651 & 12956\\\hline
		$(6,4)$ & 533 & 820 & $(11,3)$ & 328 & 1230\\\hline
		$(8,4)$ & 1599 & 3485 & $(12,2)$ & 0 & 123 \\\hline
		$(9,3)$ & 0 & 246 & $(12,4)$ & 42599 & 57195\\\hline
	\end{tabular}
\end{table}

As another example, we design a QC-LDPC code ${\cal C}_1$ lifted from the $3 \times 5$ fully-connected base graph with lifting degree $N=41$, with $g = 8$, and 
free of LETSs within the union of two rectangular regions $a \leq 10, b \leq 3$, and $a \leq 12, b \leq 2$. The exponent matrix of ${\cal C}_1$ is
\begin{equation}
\scriptsize
P_1=
\begin{bmatrix} 
0 & 0 & 0 & 0 & 0 \\
0 & 1 & 5 & 7 & 26 \\
0 & 3 & 13 & 30 & 37 \\
\end{bmatrix}
\:.
\end{equation}
This code has the same degree distribution and block length as the code designed in~\cite{ref 9}, but is free of LETS structures within a much larger region.
(See Table VI for the comparison of LETS distributions. In the table, we have only listed the classes for which at least one code has non-zero multiplicity). 

As another experiment, we construct a QC-LDPC code which has similar parameters to the well-known $(155,64)$ Tanner code~\cite{ref 12}, but has a lower error floor.
Tanner code is a cyclic lifting of the fully-connected $3\times 5$ base graph with $N=31$ and has $g=8$. A similar code was also designed in \cite{ref 8} with the goal of 
reducing the error floor by removing all $(5,3)$ LETSs and minimizing the number of $(6,4)$ LETSs in the Tanner graph of the code. 
In this case, we use the formulation of Problem (b) in Section V for the fully-connected $3 \times 5$ base graph with $N = 31$ and $g_0 = 8$. 
We consider two cases of $b_{\max} = 3$ and $b_{\max} = 2$, and are able to construct two codes that are free of LETS structures 
up to size $a_{\max} = 8$ and $a_{\max} = 10$, respectively. Then, we target the union of the LETSs within the two regions $a \leq 8, b \leq 3$, and $a \leq 10, b \leq 2$, 
and are able to design a code ${\cal C}_2$ with $N=31$ and $g=8$ that has no LETS within the targeted region. 
The exponent matrix of this code is as follows
\begin{equation}
\scriptsize
P_2=
\begin{bmatrix} 
0 & 0 & 0 & 0 & 0 \\
0 & 1 & 5 & 21 & 30 \\
0 & 3 & 13 & 6 & 20 \\
\end{bmatrix}
\:.
\end{equation}
We have presented the multiplicities of LETSs of the designed code and those of the Tanner code as well as the code designed in~\cite{ref 8} in Table VII. 
It is well-known that the most dominant trapping sets of the Tanner code are $(8,2)$ and $(10,2)$ LETSs, respectively. 
Both structures are completely removed from ${\cal C}_2$. On the other hand, for the code of \cite{ref 8}, although 
the $(8,2)$ LETSs are removed, there are still a number of $(10,2)$ LETSs present. This code also has $31$ instances of the 
$(7,3)$ LETS, another potentially problematic structure. We have compared the frame error rate (FER) of the constructed code 
${\cal C}_2$ with those of Tanner code and the code of~\cite{ref 8} in Fig. 3.
The superior performance of ${\cal C}_2$ over both Tanner code and the code of \cite{ref 8} in the error floor region can 
be observed. Based on the LETS multiplicities presented in Table VII, we expect the performance gap between ${\cal C}_2$ 
and the code of~\cite{ref 8} to increase by further increase in the signal-to-noise ratio (SNR). In fact, 
based on the simulation results, at SNR = $6$ dB, $44$ out of $100$ errors of the code designed in \cite{ref 8} 
are $(10,2)$ LETSs, which are completely absent in ${\cal C}_2$.   

\begin{table}[]
	\centering
	\tiny
	\caption{Multiplicities of LETS structures in the  range $a \leq 12$ and $b \leq 3$ for ${\cal C}_2$, $(155,64)$ Tanner code, and the code of~\cite{ref 8}}
	\label{my-label}
	\begin{tabular}{||c|c|c|c||c|c|c|c||}
		\hline
		$(a,b)$ class & Tanner Code & Code of \cite{ref 8} & ${\cal C}_2$ & $(a,b)$ class & Tanner Code & Code of \cite{ref 8} & ${\cal C}_2$ \\ \hline		
		$(4,4)$ & 465 & 527 & 558 & $(9,3)$ & 1860 & 558 & 465\\ \hline
		$(5,3)$ & 155 & 0 & 0 & $(10,2)$ & 1395 & 93 & 0\\ \hline
		$(7,3)$ & 930 & 31 & 0 & $(11,3)$ & 6200 & 4960 & 4154\\ \hline
		$(8,2)$ & 465 & 0 & 0 & $(12,2)$ & 930 & 992 & 682\\ \hline
	\end{tabular}
\end{table}

\begin{figure*}
	\centering \scalebox{0.60}
	{\includegraphics{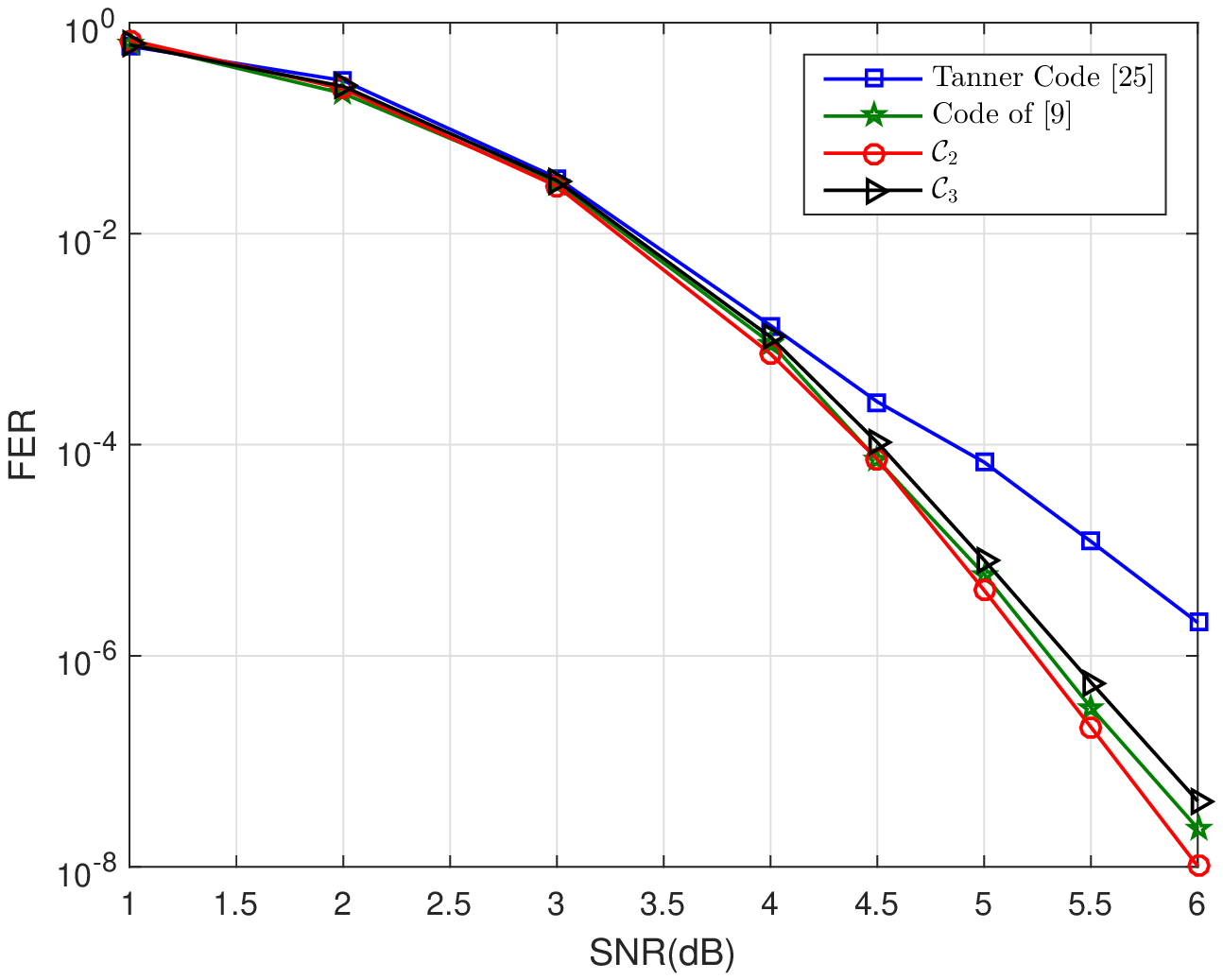}}\\\vspace{.1cm}
	{\normalsize\textbf{\footnotesize Fig. 3}\,\,\,\footnotesize FER comparison among constructed codes ${\cal C}_2$, ${\cal C}_3$, $(155,64)$ Tanner code, and the code of~\cite{ref 8} (All codes are QC,  and have $d_v=3$, $d_c=5$. 
	The code ${\cal C}_2$, $(155,64)$ Tanner code, and the code of~\cite{ref 8} have $g=8$ and block length $155$, while ${\cal C}_3$ has $g=6$ and block length $145$)}\vspace{-1.0cm}
\end{figure*}

As another example, we construct girth-$6$ codes following the formulation of Problem (a), and by using the fully-connected  $3\times 5$ base graph 
with the constraint that the code is free of LETSs within $4$ different ranges as shown in Table VIII. 
The smallest lifting degree and the exponent matrix for each case are also given in Table VIII. For the first range, $a \leq 5$ and $b \leq 2$, 
our search result is exhaustive and $N=10$ is in fact the smallest possible lifting degree that can satisfy the LETS constraint.
For the other ranges, the search is not exhaustive and the given value of $N$ provides an upper bound on the smallest lifting degree.
We have included the FER of the code ${\cal C}_3$ designed to be free of LETSs in the range $a \leq 11,b \leq 2$, in Fig. 3. 
As can be seen, this code handily outperforms the Tanner code in the error floor region. This is impressive, considering that both the 
girth and the block length of ${\cal C}_3$ are smaller than those of the Tanner code ($6$ vs. $8$, and $145$ vs. $155$, respectively).

\begin{table}[]
	\centering
	\tiny
	\caption{Upper bounds on the lifting degree of girth-$6$ QC-LDPC codes with fully-connected $3 \times 5$ base graph with no 
         LETSs within different ranges, and the corresponding exponent matrices}
	\label{my-label}
	\begin{tabular}{||c|c|c|c|c||}
		\hline
		Range & $(a \leq 5,b \leq 2)$ & $(a \leq 7,b \leq 2)$ & $(a \leq 9,b \leq 2)$ & $(a \leq 11,b \leq 2)$ \\ \hline
		$N$ & 10 & 15 & 22 & 29 \\ \hline
		Exponent Matrix & $\begin{bmatrix} 0 & 0 & 0 & 0 & 0 \\ 0 & 3 & 5 & 6 & 8 \\ 0 & 2 & 8 & 4 & 9 \\ \end{bmatrix}$ & $\begin{bmatrix} 0 & 0 & 0 & 0 & 0 \\ 0 & 2 & 7 & 10 & 14 \\ 0 & 12 & 11 & 2 & 13 \\ \end{bmatrix}$ & $\begin{bmatrix} 0 & 0 & 0 & 0 & 0 \\ 0 & 10 & 12 & 13 & 18 \\ 0 & 21 & 19 & 14 & 20 \\ \end{bmatrix}$ & $\begin{bmatrix} 0 & 0 & 0 & 0 & 0 \\ 0 & 4 & 9 & 15 & 16 \\ 0 & 8 & 16 & 1 & 18 \\ \end{bmatrix}$ \\ \hline
	\end{tabular}	
\end{table}

To further demonstrate the strength of the designed codes, we consider the code whose exponent matrix is given in the last column of Table V, and compare
it with similar codes ($3 \times 6$ fully-connected base graph, $N=80$ and $g=8$) constructed using the well-known QC-PEG~\cite{ref 2} and Improved QC-PEG \cite{ref 8} methods. 
The multiplicities of LETSs for the three codes are presented in Table IX. As can be seen from the table, many of LETSs that dominate the error floor performance of the other two codes
are absent from the Tanner graph of the designed code. To investigate this further, in Fig. 4, we have provided the FER of the three codes. 
Fig. 4 shows the superior error floor performance of the designed code. Based on the simulation results, the dominant trapping set structures of QC-PEG code
are $(8,2)$ and $(10,2)$. In the code of \cite{ref 8}, the $(8,2)$ structure has been removed and thus the error floor performance has improved compared to the QC-PEG code. 
The dominant trapping sets of the code of \cite{ref 8} are in $(10,2)$ and $(12,2)$ classes, followed by $(7,3)$ and $(9,3)$ classes. 
All of these classes however, are absent from our designed code, thus the superior error floor performance.

\begin{figure}
	\centering \scalebox{0.60}
	{\includegraphics{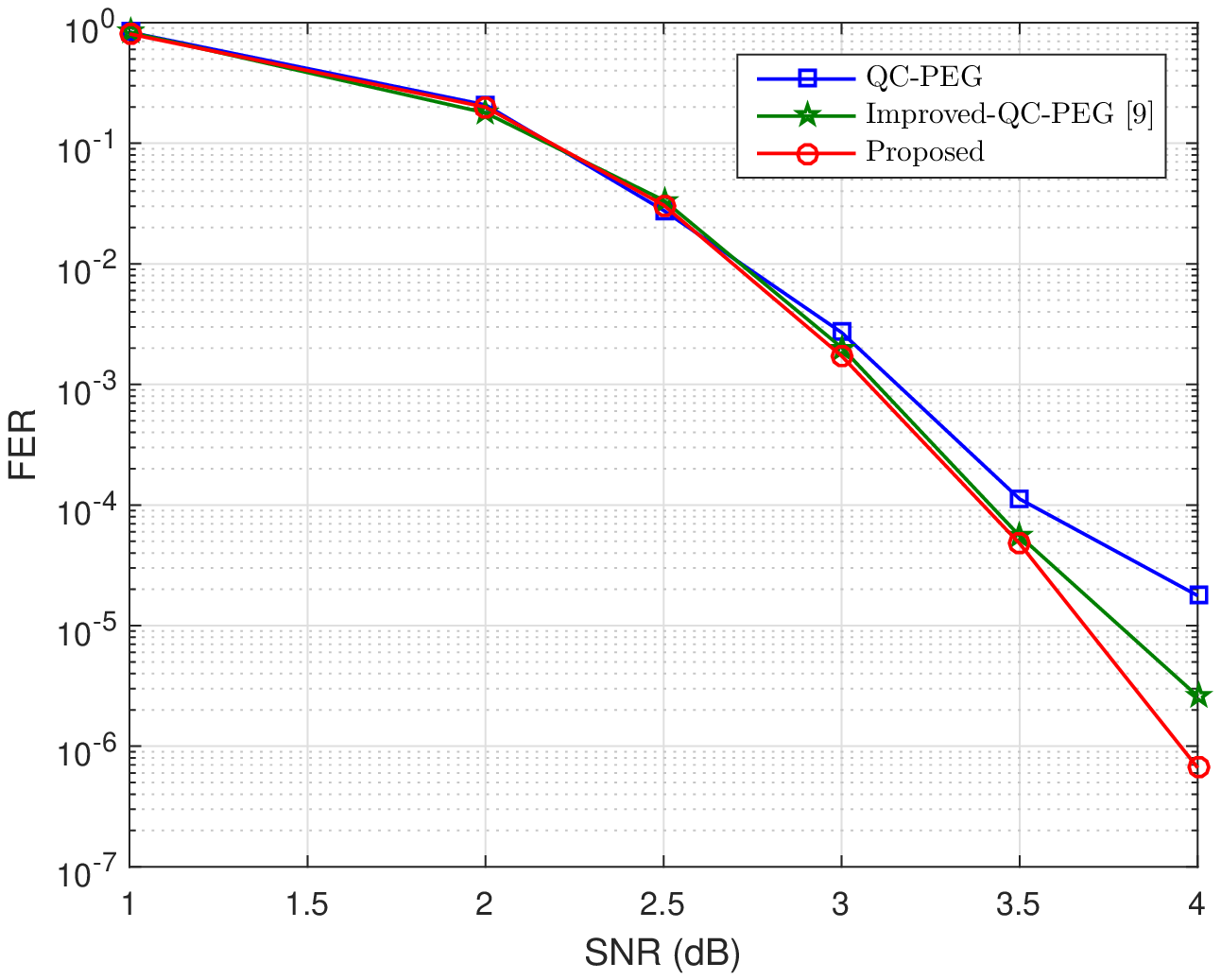}}\\\vspace{.1cm}
	{\normalsize\textbf{\footnotesize Fig. 4}\,\,\,\footnotesize FER performance of the constructed $(480,240)$ QC-LDPC code (last column of Table V) in comparison with the FER of similar codes constructed by QC-PEG \cite{ref 2} and Improved QC-PEG \cite{ref 8}}\vspace{-0.5cm}
\end{figure}

\begin{table}[]
	\centering
	\tiny
	\caption{Multiplicities of LETS structures in the range $a \leq 12$ and $b \leq 3$, for the designed code (last column of Table V) and similar codes designed by QC-PEG methods}
	\label{my-label}
	\begin{tabular}{||c|c|c|c||}
		\hline
		$(a,b)$ class & QC-PEG \cite{ref 2} & Improved QC-PEG \cite{ref 8} & Designed Code \\ \hline
		$(7,3)$ & 160 & 160 & 0 \\ \hline
		$(8,2)$ & 80 & 0 & 0 \\ \hline
		$(9,3)$ & 160 & 480 & 0 \\ \hline
		$(10,2)$ & 160 & 80 & 0 \\ \hline
		$(11,3)$ & 1280 & 1120 & 0 \\ \hline
		$(12,2)$ & 240 & 160 & 0 \\ \hline
	\end{tabular}
	\vspace{-1.0cm}
\end{table}

All the examples given so far, similar to the majority of the results available in the literature, were for codes with $d_v=3$. 
To demonstrate the generality of our method, we also construct girth-$6$ codes 
from the fully-connected $4 \times 6$, $4 \times 8$ and $4 \times 16$ base graphs following the formulation of 
Problem (a) in Section V. For each base graph, four codes are designed, where the LETS structures within $4$ different ranges are avoided. 
The results are presented in Tables X, XI, and XII, respectively. All the values of $N$ in these tables are upper bounds on 
the smallest lifting degree that can satisfy the corresponding LETS constraint, with the exception of the result of $N=7$ in Table X, 
which is in fact the smallest lifting degree that can result in a code free of LETSs within the range $a \leq 5, b \leq 5$. 
We note that Diouf \cite{ref 14} has constructed a protograph-based QC-LDPC code with $g=6$ 
and free of the $(4,4)$ LETS structure using the $4 \times 6$ fully-connected base graph with $N=7$. 
An examination of the code of \cite{ref 14} reveals that its LETS distribution within the range $a \leq 8, b \leq 5$ is the same as that of the code designed here. 

As another example, we consider the regular $(576,432)$ QC-LDPC code with $d_v=4$ designed in \cite{ref 15}. This code is designed based on array dispersion method, 
where first a $4 \times 36$ exponent matrix with lifting degree $N=36$ is constructed, and then, $16$ columns of this matrix are selected 
such that the resulting code contains fewer short cycles and larger girth. For comparison, we design a cyclic lifting ${\cal C}_4$ of the $4 \times 16$ base graph with $N=36$ and $g=6$
and free of LETSs within the union of the ranges $a \leq 5, b \leq 5$ and $a \leq 8, b \leq 3$:
\setcounter{MaxMatrixCols}{16}
\begin{equation}
\footnotesize
P_4=
\begin{bmatrix} 
0 & 0 & 0 & 0 & 0 & 0 & 0 & 0 & 0 & 0 & 0 & 0 & 0 & 0 & 0 & 0 \\
0 & 2 & 4 & 5 & 8 & 10 & 11 & 12 & 16 & 18 & 20 & 22 & 23 & 28 & 29 & 33 \\
0 & 34 & 22 & 6 & 25 & 20 & 30 & 23 & 32 & 5 & 35 & 28 & 21 & 31 & 7 & 1 \\
0 & 33 & 17 & 2 & 26 & 8 & 20 & 4 & 10 & 35 & 19 & 32 & 31 & 3 & 14 & 29
\end{bmatrix}
\:.
\end{equation}

In Table XIII, we have listed the LETSs of ${\cal C}_4$ and the code of \cite{ref 15} in the range $a \leq 8, b \leq 6$. 
The FER curves for the two codes are also given in Fig. 5. These results clearly show the superior LETS distribution and error floor performance of ${\cal C}_4$.

\begin{table}[]
	\centering
	\tiny
	\caption{Upper bounds on the lifting degree of girth-$6$ QC-LDPC codes with fully-connected $4 \times 6$ base graph with no 
         LETSs within different ranges, and the corresponding exponent matrices}
	\label{my-label}
	\begin{tabular}{||c|c|c|c|c||}
		\hline
		Range & $(a \leq 5,b \leq 5)$ & $(a \leq 6,b \leq 5)$ & $(a \leq 7,b \leq 5)$ & $(a \leq 8,b \leq 5)$ \\ \hline
		$N$ & 7 & 13 & 15 & 17\\ \hline
		Exponent Matrix & $\begin{bmatrix} 0 & 0 & 0 & 0 & 0 & 0 \\ 0 & 1 & 2 & 3 & 4 & 5 \\ 0 & 2 & 4 & 6 & 1 & 3 \\ 0 & 4 & 1 & 5 & 2 & 6 \end{bmatrix}$ & $\begin{bmatrix} 0 & 0 & 0 & 0 & 0 & 0 \\ 0 & 8 & 9 & 10 & 11 & 12 \\ 0 & 3 & 10 & 8 & 4 & 2 \\ 0 & 2 & 4 & 6 & 1 & 11 \end{bmatrix}$ & $\begin{bmatrix} 0 & 0 & 0 & 0 & 0 & 0 \\ 0 & 1 & 3 & 7 & 8 & 13 \\ 0 & 2 & 6 & 3 & 12 & 7 \\ 0 & 3 & 10 & 6 & 5 & 4 \end{bmatrix}$ & $\begin{bmatrix} 0 & 0 & 0 & 0 & 0 & 0 \\ 0 & 6 & 13 & 14 & 15 & 16 \\ 0 & 4 & 15 & 10 & 5 & 7 \\ 0 & 3 & 6 & 4 & 2 & 5 \end{bmatrix}$ \\ \hline		
	\end{tabular}
\end{table}

\begin{table}[]
	\centering
	\tiny
	\caption{Upper bounds on the lifting degree of girth-$6$ QC-LDPC codes with fully-connected $4 \times 8$ base graph with no 
         LETSs within different ranges, and the corresponding exponent matrices}
	\label{my-label}
	\begin{tabular}{||c|c|c||}
		\hline
		Range & $(a \leq 5,b \leq 5)$ & $(a \leq 6,b \leq 5)$ \\ \hline
		$N$ & 15 & 18 \\ \hline
		Exponent Matrix & $\begin{bmatrix} 0 & 0 & 0 & 0 & 0 & 0 & 0 & 0 \\ 0 & 4 & 5 & 9 & 11 & 12 & 13 & 14 \\ 0 & 7 & 6 & 14 & 5 & 10 & 12 & 1 \\ 0 & 14 & 3 & 8 & 13 & 9 & 1 & 6 \end{bmatrix}$ & $\begin{bmatrix} 0 & 0 & 0 & 0 & 0 & 0 & 0 & 0 \\ 0 & 9 & 10 & 13 & 14 & 15 & 16 & 17 \\ 0 & 5 & 2 & 4 & 15 & 1 & 11 & 16 \\ 0 & 12 & 7 & 1 & 9 & 17 & 2 & 4 \end{bmatrix}$ \\ \hline
		Range & $(a \leq 7,b \leq 5)$ & $(a \leq 8,b \leq 5)$ \\ \hline
		$N$ & 21 & 24 \\ \hline
		Exponent Matrix & $\begin{bmatrix} 0 & 0 & 0 & 0 & 0 & 0 & 0 & 0 \\ 0 & 4 & 15 & 16 & 17 & 18 & 19 & 20 \\ 0 & 17 & 9 & 5 & 4 & 1 & 20 & 19 \\ 0 & 6 & 5 & 4 & 15 & 3 & 14 & 12 \end{bmatrix}$ & $\begin{bmatrix} 0 & 0 & 0 & 0 & 0 & 0 & 0 & 0 \\ 0 & 11 & 15 & 17 & 19 & 20 & 21 & 23 \\ 0 & 22 & 5 & 21 & 13 & 2 & 14 & 20 \\ 0 & 10 & 9 & 18 & 7 & 16 & 6 & 13 \end{bmatrix}$ \\ \hline
	\end{tabular}
\end{table}

\setcounter{MaxMatrixCols}{16}
\begin{table}[]
	\centering
	\tiny
	\caption{Upper bounds on the lifting degree of girth-$6$ QC-LDPC codes with fully-connected $4 \times 16$ base graph with no 
         LETSs within different ranges, and the corresponding exponent matrices}
	\label{my-label}
	\begin{tabular}{||c|c||}
		\hline
		Range & Exponent Matrices \\ \hline
		
		$N=32$, $(a \leq 5,b \leq 5)$ &
		$\begin{bmatrix}
		0 & 0 & 0 & 0 & 0 & 0 & 0 & 0 & 0 & 0 & 0 & 0 & 0 & 0 & 0 & 0 \\
		0 & 2 & 3 & 4 & 6 & 10 & 12 & 14 & 16 & 18 & 20 & 23 & 24 & 25 & 28 & 31 \\ 
		0 & 17 & 16 & 25 & 20 & 3 & 29 & 22 & 11 & 6 & 27 & 2 & 8 & 23 & 15 & 5 \\ 
		0 & 12 & 21 & 2 & 8 & 25 & 18 & 7 & 10 & 13 & 31 & 30 & 9 & 16 & 27 & 3 
		\end{bmatrix}$ \\ \hline
		
		$N=44$, $(a \leq 6,b \leq 5)$ &
		$\begin{bmatrix}
		0 & 0 & 0 & 0 & 0 & 0 & 0 & 0 & 0 & 0 & 0 & 0 & 0 & 0 & 0 & 0 \\
		0 & 1 & 3 & 5 & 8 & 9 & 11 & 12 & 16 & 17 & 20 & 23 & 24 & 26 & 33 & 39 \\ 
		0 & 15 & 31 & 13 & 21 & 2 & 35 & 32 & 3 & 34 & 42 & 9 & 30 & 11 & 29 & 7 \\ 
		0 & 20 & 9 & 30 & 16 & 37 & 29 & 11 & 18 & 22 & 43 & 21 & 7 & 17 & 13 & 4 
		\end{bmatrix}$ \\ \hline
		
		$N=54$, $(a \leq 7,b \leq 5)$ &
		$\begin{bmatrix}
		0 & 0 & 0 & 0 & 0 & 0 & 0 & 0 & 0 & 0 & 0 & 0 & 0 & 0 & 0 & 0 \\
		0 & 2 & 3 & 5 & 7 & 10 & 12 & 13 & 16 & 18 & 20 & 21 & 22 & 23 & 29 & 50 \\ 
		0 & 47 & 7 & 28 & 39 & 20 & 53 & 27 & 6 & 25 & 8 & 24 & 11 & 35 & 22 & 34 \\ 
		0 & 5 & 28 & 45 & 9 & 29 & 13 & 47 & 33 & 30 & 24 & 53 & 3 & 17 & 52 & 11 
		\end{bmatrix}$ \\ \hline
		
		$N=60$, $(a \leq 8,b \leq 5)$ &
		$\begin{bmatrix}
		0 & 0 & 0 & 0 & 0 & 0 & 0 & 0 & 0 & 0 & 0 & 0 & 0 & 0 & 0 & 0 \\
		0 & 1 & 2 & 3 & 8 & 10 & 12 & 14 & 15 & 17 & 20 & 21 & 26 & 28 & 37 & 40 \\ 
		0 & 5 & 49 & 25 & 40 & 15 & 27 & 35 & 29 & 24 & 9 & 30 & 42 & 4 & 18 & 39 \\ 
		0 & 3 & 40 & 47 & 36 & 23 & 19 & 45 & 6 & 21 & 8 & 55 & 49 & 42 & 1 & 32 
		\end{bmatrix}$ \\ \hline
			
	\end{tabular}
\end{table} 

\begin{table}[]
	\centering
	\tiny
	\caption{Multiplicities of LETS structures in the range $a \leq 8$ and $b \leq 6$ for the constructed code ${\cal C}_4$ and the code of~\cite{ref 15} }
	\label{my-label}
	\begin{tabular}{||c|c|c||c|c|c||}
		\hline
		$(a,b)$ class & ${\cal C}_4$ & Code of \cite{ref 15} & $(a,b)$ class & ${\cal C}_4$ & Code of \cite{ref 15} \\ \hline
		$(3,6)$ & 14580 & 12456 & $(7,4)$ & 2340 & 20160 \\ \hline
		$(4,4)$ & 0 & 144 & $(7,6)$ & 590724 & 1122480 \\ \hline
		$(4,6)$ & 27000 & 27648 & $(8,0)$ & 0 & 171 \\ \hline
		$(5,4)$ & 0 & 324 & $(8,2)$ & 0 & 2520 \\ \hline
		$(5,6)$ & 59508 & 88416 & $(8,4)$ & 14634 & 96732 \\ \hline
		$(6,4)$ & 756 & 7236 & $(8,6)$ & 2345328 & 3616272 \\ \hline
		$(6,6)$ & 189360 & 359136 & & &  \\ \hline
	\end{tabular}
	\vspace{-0.5cm}
\end{table}

\begin{figure*}
	\centering \scalebox{0.60}
	{\includegraphics{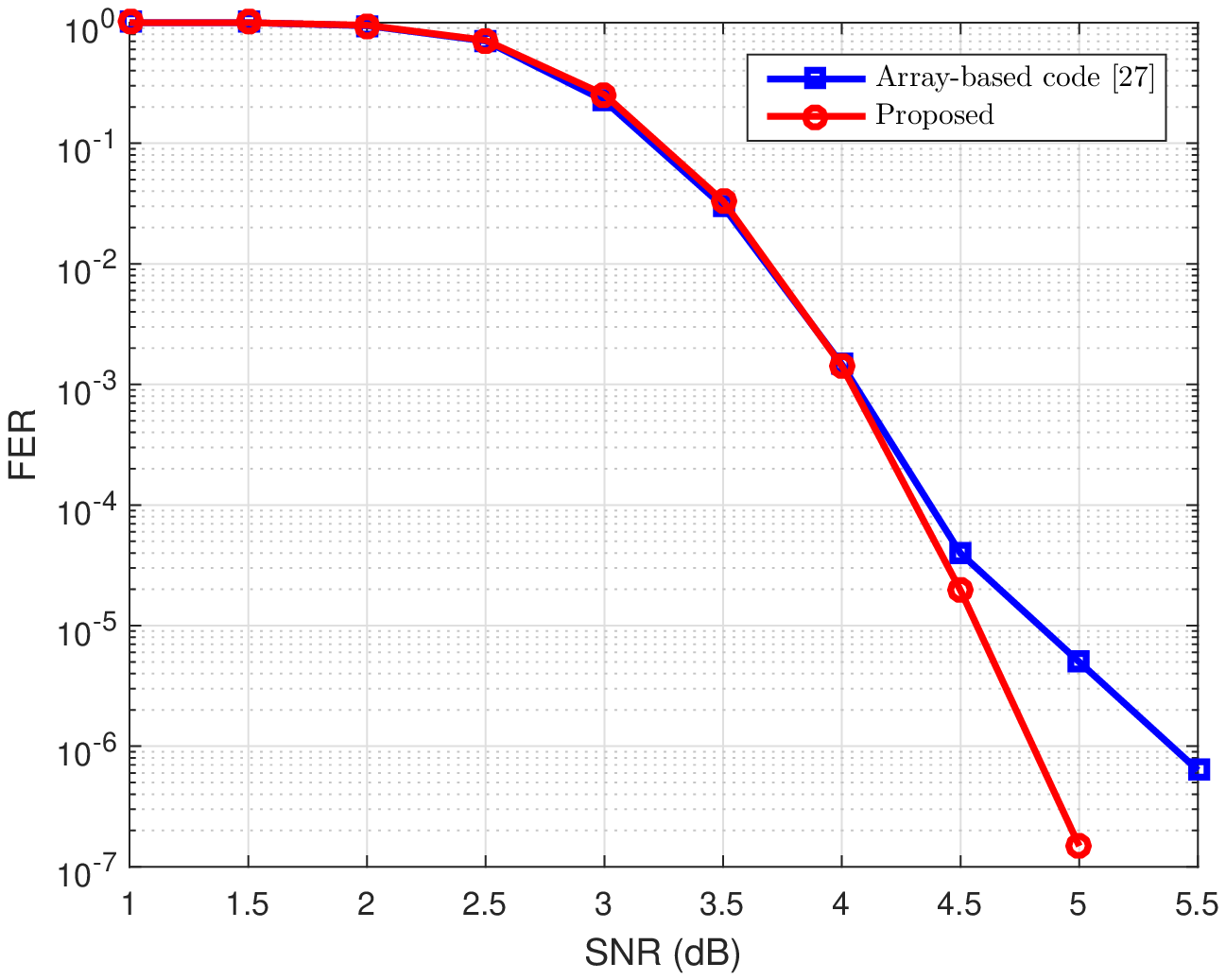}}\\\vspace{.1cm}
	{\normalsize\textbf{\footnotesize Fig. 5}\,\,\,\footnotesize FER curves of the constructed $(576,432)$ code ${\cal C}_4$ and the similar code of \cite{ref 15}} \vspace{-0.5cm}
\end{figure*}

As the final example, we consider a $(2133,1817)$ array-based code with lifting degree $N=79$, $d_v=4$, and $d_c=27$. To construct this code, similar to the previous example, 
first a $4 \times 79$ exponent matrix is generated, and then $27$ columns of this matrix are chosen such that the resulting code has fewer short cycles and larger girth~\cite{ref 15}. 
For comparison, we then use our technique to construct a code ${\cal C}_5$ with similar parameters ($N=79$, $d_v=4$, $d_c=27$) which is free of LETSs within the union of 
the ranges $a \leq 7, b \leq 5$, and $a \leq 8, b \leq 3$. The exponent matrix of ${\cal C}_5$ is given by:
\setcounter{MaxMatrixCols}{27}
\begin{equation}
\footnotesize
P_5=
\setlength{\arraycolsep}{2.4pt}
\renewcommand{\arraystretch}{0.8}
\begin{bmatrix}
0 & 0 & 0 & 0 & 0 & 0 & 0 & 0 & 0 & 0 & 0 & 0 & 0 & 0 & 0 & 0 & 0 & 0 & 0 & 0 & 0 & 0 & 0 & 0 & 0 & 0 & 0 \\
0 & 2 & 3 & 5 & 6 & 9 & 11 & 12 & 13 & 17 & 19 & 21 & 24 & 28 & 30 & 34 & 36 & 37 & 38 & 46 & 49 & 51 & 52 & 55 & 60 & 64 & 69 \\
0 & 24 & 68 & 66 & 36 & 59 & 37 & 45 & 29 & 58 & 64 & 75 & 34 & 2 & 57 & 70 & 55 & 35 & 40 & 27 & 1 & 11 & 67 & 72 & 65 & 23 & 32 \\
0 & 4 & 12 & 76 & 43 & 53 & 8 & 54 & 34 & 66 & 22 & 77 & 72 & 55 & 36 & 35 & 15 & 25 & 13 & 41 & 62 & 68 & 56 & 78 & 10 & 38 & 9
\end{bmatrix}
\:.
\end{equation}

The exhaustive search of LETSs within the range of $a \leq 8$ and $b \leq 5$ for ${\cal C}_5$ reveals that this code has only one class, i.e., $(8,4)$, within this range with non-zero multiplicity. The multiplicity of $(8,4)$ LETSs in ${\cal C}_5$ is $5925$. On the other hand, the code of \cite{ref 15} has three classes with non-zero multiplicity in the range $a \leq 8$, $b \leq 5$. These classes are $(7,4), (8,2)$ and $(8,4)$, with multiplicities $9006, 3634$, and $122450$, respectively.  As can be seen, the designed code has a superior LETS distribution, and thus a lower error floor, compared to the code of \cite{ref 15}. In fact, our simulations show that the most dominant class of trapping sets in  the code of \cite{ref 15} is the $(8,2)$ class, which is completely absent in the designed code.

Finally, in order to demonstrate the complexity reduction of the proposed search technique in comparison with the $dpl$ search of~\cite{ref 10}, in Table~XIV, we have listed the run-time of both algorithms for finding the solutions in the largest ranges of Tables~IV, V, X, XI, and XII. As can be seen, in all cases, the proposed algorithm is much faster than that of  \cite{ref 10} by up to 
more than one order of magnitude.

\begin{table}[]
	\centering
	\scriptsize
	\caption{Comparison of run-times for the proposed method and the method of \cite{ref 10}}
	\label{my-label}
	\begin{tabular}{||c|c|c|c|c|c||}
		\hline
		Range of $a$ and $b$ values & $(a \leq 12,b \leq 3)$ & $(a \leq 12,b \leq 3)$ & $(a \leq 8,b \leq 5)$ & $(a \leq 8,b \leq 5)$ & $(a \leq 8,b \leq 5)$ \\ \hline
		Girth $g$ & $8$ & $8$ & $6$ & $6$ & $6$ \\ \hline
     	Variable node degree $d_v$ & $3$ & $3$ & $4$ & $4$ & $4$ \\ \hline
     	Check node degree $d_c$ & $5$ & $6$ & $6$ & $8$ & $16$ \\ \hline  
		Lifting degree $N$ & $46$ & $80$ & $17$ & $24$ & $60$ \\ \hline
		Run-time of the proposed method (sec.) & $2889$  & $3425$  & $755$  & $2642$  & $487$  \\ \hline
		Run-time of the method of \cite{ref 10} (sec.) & $36293$  & $42955$  & $5518$  & $14718$  & $3591$ \\ \hline		
	\end{tabular}
	\vspace{-0.5cm}
\end{table}

\section{Conclusion}
In this paper, we proposed a systematic and efficient method to construct protograph-based QC-LDPC codes. We first examined 
the trapping set structures of such codes, and demonstrated that some of the structures that can exist in a general (randomly constructed) code
cannot exist in codes that have QC structure. This was done through the transformation of the problem into a graph coloring problem and was 
the first step in a series of steps devised to simplify the design of QC-LDPC codes. The next step was to develop an efficient layered $dpl$-based 
search algorithm for finding a targeted set of trapping sets that are to be avoided in the code (for a good error floor performance).
The algorithm was devised as a backward recursion to minimize the number of intermediate structures and the expansions that were needed to search for the targeted trapping sets,
as well as to use structures that have a higher chance of having a lower multiplicity in the graph. Numerous codes were constructed using the proposed 
design technique with superior performance compared to the existing codes in the literature. The systematic 
approach of the design makes it applicable to codes with different node degrees (rate), girths and block lengths with the flexibility of selecting 
any set of target trapping sets. The efficiency of the search algorithm makes it possible to design codes with larger degrees and block lengths which are 
free of trapping sets in larger regions compared to what was achievable before.    

We note that while our discussions in this paper are limited to QC-LDPC codes with fully-connected base graphs, many of the ideas can also be applied, in principle,
to the construction of other types of QC-LDPC codes including those with irregular or partially-connected base graphs. The details of such constructions can be an interesting 
topic of future research. We also note that the proposed search algorithm of Section IV can be used in combination with any design technique that involves 
searching a Tanner graph for certain targeted trapping set structures throughout the construction process. The construction technique of Section V is just one such example.
Moreover, the layered characterization/search algorithm of LETSs, proposed in Section IV, can also be used to construct LDPC codes with low error
floor that lack the QC structure. The only difference, compared to what is presented in this paper for QC-LDPC codes, is that for codes lacking the QC structure 
the number of possible LETS structures is larger.

\end{document}